\documentclass[a4paper,fleqn,usenatbib]{mnras}

\usepackage{newtxtext,newtxmath}

\usepackage[T1]{fontenc}
\usepackage{ae,aecompl}

\usepackage{graphicx}	
\usepackage{amsmath}	
\usepackage{amssymb}	
\usepackage[shortlabels]{enumitem}      
\usepackage{svg}    
\usepackage{booktabs}   
\usepackage[nice]{nicefrac}  

\newcommand{\red}[1]{{#1}}

\newcommand{\dtg}{\ensuremath{\Delta t_\text{gas}}}
\newcommand{\dtdm}{\ensuremath{\Delta t_\text{DM}}}

\graphicspath{{Figures}}

\title[DMAF in Cosmological Simulations]{A Novel Scheme for Dark Matter Annihilation Feedback in Cosmological Simulations}

\author[List et al.]{
Florian List$^{1}$\thanks{E-mail: flis0155@uni.sydney.edu.au},
Nikolas Iwanus$^{1}$,
Pascal J. Elahi$^{2,3}$, and
Geraint F. Lewis$^{1}$
\\
$^{1}$Sydney Institute for Astronomy, School of Physics, A28, The University of Sydney, NSW 2006, Australia\\
$^{2}$International Centre for Radio Astronomy Research, University of Western Australia, 35 Stirling Highway, Crawley, WA 6009, Australia \\
$^{3}$ARC Centre of Excellence for All Sky Astrophysics in 3 Dimensions (ASTRO 3D)
}

\date{Accepted XXX. Received YYY; in original form ZZZ}

\pubyear{2019}

\begin{document}
\label{firstpage}
\pagerange{\pageref{firstpage}--\pageref{lastpage}}
\maketitle

\begin{abstract}
We present a new self-consistent method for incorporating dark matter annihilation feedback (DMAF) in cosmological N-body simulations. The power generated by DMAF is evaluated at each dark matter (DM) particle which allows for flexible energy injection into the surrounding gas based on the specific DM annihilation model under consideration. Adaptive, individual time steps for gas and DM particles are supported \red{and a new time-step limiter, derived from the propagation of a Sedov--Taylor blast wave, is introduced}. We compare this donor-based approach with a receiver-based approach used in recent studies \red{and illustrate the differences by means of a toy example. Furthermore, we consider an isolated halo and a cosmological simulation and show that for these realistic cases, both methods agree well with each other.} The extension of our implementation to scenarios such as non-local energy injection, velocity-dependent annihilation cross-sections, and DM decay is straightforward.
\end{abstract}




\section{Introduction}
The history of cosmological N-body simulations (in an admittedly broad sense) dates back to 1941 when \citet{Holmberg1941} conducted light bulb experiments in order to investigate the movement of galaxies resulting from gravitational interactions. Since the dawn of N-body simulations by means of computers in the early 1960's \citep{vonHoerner1960, Aarseth1963} where the particle number was limited to $N \leq 100$, technical innovation has led the increase of computational power to keep up with Moore's law and presently, simulations with many billions of particles shed light onto the non-linear growth of structure \red{(see e.g. \citealt{Springel2005, Springel2008, Springel2017, Boylan-Kolchin2009, Stadel2009, Klypin2011, Vogelsberger2014, Schaye2015}} for high-resolution simulations performed in the last two decades). 
\par While cosmological simulations arguably constitute a major pillar in the establishment of $\Lambda$ cold dark matter ($\Lambda \text{CDM}$) as the concordance model of cosmology and confirm its predictions on a large-scale, tensions exist on smaller scales. Examples are the missing satellite problem that describes the overabundance of satellite galaxies predicted by $\Lambda \text{CDM}$ simulations as compared to observations of the Milky Way \citep{Klypin1999, Moore1999} and the Local Group \citep{Zavala2009} and the core-cusp problem \citep{Moore1994, DeBlok2001} that addresses the contrast between cuspy halo density profiles predicted by $\Lambda \text{CDM}$ and flattened DM cores as observed in low surface brightness galaxies, amongst other discrepancies. The entirety of differences between $\Lambda \text{CDM}$ and observations is sometimes called the ``small-scale crisis'' (see \citealt{Weinberg2015} for a review). 
\par Incorporating baryonic physics into cosmological simulations may alleviate the small-scale discrepancies, e.g. by suppressing galaxy formation due to photoionization \citep{Bullock2000, Benson2002, Somerville2002}, supernova feedback \citep{Governato2012, Pontzen2012}, or tidal stripping \citep{Zolotov2012, Brooks2013}. Another avenue of investigation is to question the cold and collisionless nature of DM: hypothesising ``warm'' DM with a non-negligible velocity in the early Universe instead of the CDM paradigm leads to a sharp cut-off in the high-frequent modes of the power spectrum and hence to the suppression of small-scale structure below the free-streaming scale \citep{Lovell2012}. DM particles that strongly interact with each other, known as self-interacting DM, may also reconcile predictions from simulations with observations \red{\citep{Vogelsberger2012, Zavala2013, Rocha2013, Peter2013, Elbert2015, Tulin2018}}, although this is disputed \citep{KuzioDeNaray2010, Schneider2014}. A further possible explanation for the erasure of small-scale structure is late kinetic decoupling of a cold DM species that remains in thermal equilibrium with a relativistic species until the Universe has cooled down to sub-$\text{keV}$ temperatures \citep{Boehm2014, Bringmann2016}. 
\par In the course of the last decades, the tools for cosmological simulations have developed from purely gravitational codes to complex programs which model gas hydrodynamics and additional baryonic physics such as stellar feedback, cosmic rays, supernovae, feedback from active galactic nuclei, and various gas cooling and heating mechanisms. In contrast, a versatile toolbox for the investigation of the most abundant matter component, namely DM, is often lacking. In this work, we present an extension of the simulation code \textsc{Gizmo} \citep{GIZMO} to a generic class of DM particles that annihilate into standard model (SM) particles and deposit a fraction of the released energy into the intergalactic medium.
This includes in particular weakly interacting massive particles (WIMPs) which are (still, despite non-detection hitherto) one of the most promising DM candidates. WIMPs were in a thermal equilibrium with the plasma in the early Universe and the WIMP abundance was kept in an equilibrium by annihilation into and production from SM particles. As the Universe gradually expanded, the reactions of the WIMPs became too rare to maintain the equilibrium abundance, a moment called the ``freeze-out''. Calculating the velocity cross-section needed to explain the observed DM density today for a WIMP yields a velocity cross-section of $\langle \sigma v \rangle \sim 3 \times 10^{-26} \ \text{cm}^3 \ \text{s}^{-1}$ \citep{Steigman2012}, which falls into the weak scale where many well-motivated particles in extensions of the SM reside, such as the lightest supersymmetric particle.
\par The DM annihilation alters the thermal history of the Universe, leaving its imprint on the cosmic microwave background (CMB) anisotropies (e.g. \citealt{Slatyer2009}), for which reason CMB data from the Planck satellite provides tight constraints on dark matter candidates \citep[see Figure 42]{PlanckCollaboration2018} with energies $\mathcal{O}(\text{few GeV})$, assuming an s-wave annihilation channel, a thermal relic cross-section, and $2 \to 2$ annihilation.
\par Since the annihilation of DM particles at a mass scale of $\gtrsim \text{GeV}$ is expected to produce $\gamma$-rays which could be detected, further constraints come from indirect dark matter searches such as Fermi-LAT, HESS, HAWC, Chandra, and AMS-02, amongst others (e.g. \citealt{Hoof2018} for a recent review of DM signals from Fermi-LAT). In the years to come, the parameter space for WIMPs will be further narrowed down aiming to cover the entire mass range from $\sim 100 \ \text{keV}$ (below, WIMPs no longer behave like CDM) to the unitarity bound of $\sim 100 \ \text{TeV}$ \citep{Smirnov2019} using direct detection at colliders such as the LHC \citep{Abdallah2015}, underground detectors (see \citealt{Schumann2019}), and indirect searches; in particular, measurements from the LSST will enable precise probes for theoretical DM models \citep{LSSTDarkMatterGroup2019}.
\par Whereas the DM mass is highly constrained for specific annihilation channels, the strongest model-independent lower bound for s-wave $2 \to 2$ annihilation is only $\gtrsim 20 \ \text{GeV}$ \citep{Leane2018}, and for more complicated annihilation channels such as a dark annihilation products, the DM mass is even less constrained. 
\par Recently, the debate on the source of the Galactic Centre Excess (GCE), an excess in GeV $\gamma$-ray emission that could possibly be explained by faint millisecond pulsars \citep{Bartels2016, Lee2016, Macias2018}, has flared up again as \citet{Leane2019} showed that a DM annihilation signal could be mistakenly attributed to point sources due to mismodelling and thus remains a viable explanation.
The same DM candidate that might perhaps cause the GCE could also accommodate observations by the AMS-02 collaboration of an excess of $\sim 10-20 \ \text{GeV}$ cosmic-ray antiprotons \citep{Cholis2019, Cuoco2019}. \red{A $\gamma$-ray signal from the atypical globular cluster Omega Centauri, which is conjectured to be the remnant core of a stripped dwarf galaxy (e.g. \citealt{Ibata2019}), might also originate from annihilating DM \citep{Brown2019, Reynoso-Cordova2019}.}
\par For the distinction between astrophysical sources and the effects of DM annihilation, numerical simulations are a powerful tool that allows to discern the different physics involved by simply switching on and off the respective heating mechanisms. Besides, cosmological simulations are predestined for analysing the imprint of DM annihilation on structure formation, 
such as the delayed formation of galaxies \citep{Schon2015, Schon2017}.
\par The results of N-body simulations have been used to estimate annihilation fluxes \citep{Stoehr2003} and to investigate their experimental detectability \citep{Pieri2011}. \citet{Natarajan2008} coupled a galaxy model to the ``Millennium simulation'' \citep{Springel2005} to explore the energy production of DMAF during galaxy formation, and \citet{Smith2012} modelled DMAF during the collapse of primordial minihalos using an analytical DM density profile in an N-body simulation. A model for DM decay into dark radiation is considered in \citet{Dakin2019}. 
\par This work builds on \citet{Iwanus2017}, where a method for DMAF into SM particles was presented that self-consistently incorporates the DMAF power generated by the DM particles in the course of the cosmological simulation instead of resorting to an analytic model. Whereas the authors of that work propose to calculate the DMAF power at each gas N-body particle, we present a new method herein in which the DMAF power is determined at each DM N-body particle and then injected into the surrounding gas. This permits the choice of various injection mechanisms that take into account the mean free path of a specific annihilation channel.  
\par The paper is structured as follows: in Section \ref{sec:DMAF}, we give a brief overview of DM annihilation and the resulting generation of energy. Then, we introduce our numerical method in Section \ref{sec:method}, focusing on different injection schemes and on the individual time step scheme. Section \ref{sec:comparison} is dedicated to a juxtaposition between our method and the receiver-based method proposed in \citet{Iwanus2017}. Numerical results are presented in Section \ref{sec:results}, and we conclude the paper in Section \ref{sec:conclusions}.

\section{Dark matter annihilation}
\label{sec:DMAF}
We briefly summarise the fundamental equations for the pair annihilation of DM which describe the mass loss and the energy deposition into the surrounding gas.
For the sake of simplicity, we present the formulae for Majorana particles here; the only difference in case of Dirac particles is an additional correction factor of $1/2$ \citep{Abdallah2015}.
\par For DM pair annihilation, the decrease in number density is given by
\begin{equation}
\frac{dn_\chi}{dt}=-\langle\sigma v\rangle n_{\chi}^2,
\end{equation}
where $n_\chi$ is the number density of DM with respect to a comoving volume and $\langle \sigma v \rangle$ is the velocity-averaged annihilation cross-section. 
Note that the annihilation rate is $\frac{1}{2} \langle\sigma v\rangle n_{\chi}^2$, with each annihilation removing two DM particles.
\par Consequently, the mass loss due to DM annihilation within a fixed volume is
\begin{equation}
\frac{dM_\chi}{dt} = -\frac{\langle\sigma v\rangle}{m_\chi} \rho_\chi M_\chi,
\end{equation}
where $m_\chi$ denotes the DM particle mass, $\rho_\chi = m_\chi n_\chi$ is the DM mass density, and $M_\chi$ stands for the total DM mass enclosed in the selected volume.
\subsection{DMAF energy rate}
By energy conservation, each pair annihilation releases an energy of $2 m_\chi c^2$. Depending on the mean free path of the annihilation products, this energy is injected directly into the adjacent gas (typically within a few kpcs for an electron-positron decay channel due to inverse Compton scattering, \citealt{Delahaye2010}), gradually released over a larger mean free path in case of photon production, or it escapes the local vicinity as e.g. in case of a neutrino decay channel.
\par Thus, the energy rate absorbed by the surrounding gas can be written as
\begin{equation}
\frac{dE}{dt} = B f \frac{\langle\sigma v\rangle}{m_\chi} \rho_\chi M_\chi c^2,
\label{eq:energy_rate}
\end{equation}
where $B$ is a boost factor that accounts for unresolved clumps of DM (see \citealt{Bergstrom1999}), which recent studies find to be as high as $\mathcal{O}(10)$ \citep{Stref2017, Hiroshima2018} in certain situations such as for large clusters at low redshift. Additionally, $f$ is the fraction of energy that is injected into the surrounding gas, which is generally in the range of 0.01 -- 1 (see \citealt{Slatyer2009, Galli2013, Madhavacheril2014} for detailed calculations with dependence on redshift and annihilation channel).
Henceforth, we will assume $B = f = 1$ for simplicity, but in principle, arbitrary boost factors $B$ and absorption fractions $f$ can be considered with our method, such as e.g. a boost factor that depends on the local DM density \citep{Ascasibar2007}.

\section{Numerical method}
\label{sec:method}
In this section, we will present our DMAF implementation into the cosmological simulation code \textsc{Gizmo} \citep{GIZMO}, which is an offspring of the \textsc{Gadget} series of simulation codes \citep{Gadget1, Gadget2}. \textsc{Gizmo} is a parallel multi-physics code that comes with a broad range of physics modules for e.g. cooling, supernovae, cosmic rays, black halo physics, etc. \citep{FIRE2}, which can be individually enabled or disabled as one requires. In this spirit, we implemented DMAF in a modular way allowing for the simple combination with other physics modules. In contrast to older cosmological simulation codes, \textsc{Gizmo} features not only traditional smoothed-particle hydrodynamics (SPH), but also more advanced methods such as pressure SPH, meshless finite mass method, and meshless finite volume method.\footnote{For the meshless methods, the spatial discretisation uses cells rather than SPH ``particles'', but we will continue using the notion of particles for intuition. Henceforth, a ``particle'' will refer to an N-body particle and \emph{not} to a physical particle candidate if not stated otherwise.} The underlying equation for the fluid dynamics is the Euler equation, although extensions such as the Navier--Stokes equation can be enabled. Self-gravity is treated with a hybrid TreePM scheme, and the time stepping allows for distinct, adaptive time steps for individual particles.
\par In order to implement equation \eqref{eq:energy_rate} into \textsc{Gizmo}, we calculate the DM density at each DM particle. The velocity cross-section is taken to be the standard thermal relic cross-section herein, but the extension to velocity-dependent cross-sections modelling p-wave annihilation does not pose a difficulty since the velocity of each DM particle is computed anyway. For incorporating Sommerfeld enhancement \citep{Sommerfeld1931}, relative velocities between DM particles would need to be calculated in addition. Furthermore, simulating decay of DM in lieu of annihilation can easily be accomplished by modifying the implementation of equation \eqref{eq:energy_i_j}. 
\par A neighbour search at each DM particle is required to find the gas neighbours that will absorb the generated energy. As suggested in \citet{HopkinsSupernovae} in the context of supernova feedback, we carry out a bidirectional search in order to include the gas particles within the search radius of a DM particles (which is itself determined by imposing a fixed number of desired gas neighbours), as well as gas particles containing the injecting DM particle within \emph{their} kernel. This prevents us from neglecting gas particles in less dense regions and thus from introducing an unnatural bias towards high-density regions.
\par The energy generated by a DM particle $i$ of mass $M_i$ that is released into a specific gas particle $j$ found in the neighbour search is
\begin{equation}
\frac{dE_{i \to j}}{dt} = \frac{\langle\sigma v\rangle}{m_\chi} \rho_{\chi, i} M_i c^2 \frac{w_j}{\sum_{k \in \mathcal{N}_\text{gas}(i)} w_k},
\label{eq:energy_i_j}
\end{equation}
where $w_k$ are weights that specify the fraction of the entire energy produced by DM particle $i$ injected into gas particle $k$, $\mathcal{N}_\text{gas}(i)$ is the set of gas neighbours of particle $i$, and $\rho_{\chi, i}$ is the DM density evaluated at particle $i$. By normalising with the sum of all weights, the total energy received by the gas particles equals the energy produced at the DM particle.
\subsection{Choosing the weights}
\label{subsec:weights}
\begin{figure}
  \centering
  \noindent
  \resizebox{\columnwidth}{!}{
  \includegraphics{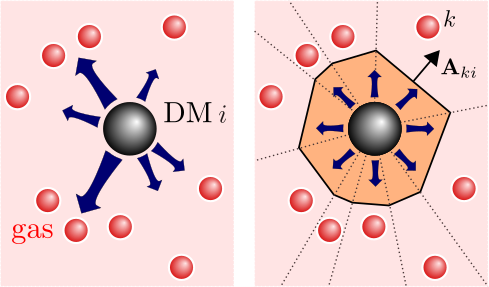}
  }
  \caption{Scheme of the energy injection from one DM particle $i$ into the surrounding gas for the choice of weights a) (\emph{left}) and b) (\emph{right}). For case b), the orange area depicts the convex hull around particle $i$. While the energy is distributed anisotropically for case a) dependent on the distribution of the surrounding gas, the energy injection is (approximately) statistically isotropic (see \citealt{HopkinsSupernovae} for an numerical investigation of this property).} 
  \label{fig:Injection_schemes}
\end{figure}
Equation \eqref{eq:energy_i_j} provides the flexibility to customise the energy injection to model a wide range of annihilation mechanisms and decay channels by defining the weights $w_k$ appropriately. In this work, we consider two possible choices:
\begin{enumerate}[a), align=left]
\item mass weighted injection:
\begin{equation}
w_k = M_k W(r_{k i}, h_i),
\end{equation}
\item solid angle weighted injection: 
\begin{equation}
w_k = \frac{1}{2}\left(1-\left(1+\left(\mathbf{A}_{k i} \cdot \hat{\mathbf{r}}_{k i}\right) /\left(\pi\left|\mathbf{r}_{k i}\right|^{2}\right)\right)^{-\nicefrac{1}{2}}\right).
\end{equation}
\end{enumerate}
\par For the mass weighted injection in case a), $W = W(r_{k i}, h_i)$ is the kernel function which depends on the distance $r_{k i} = |\mathbf{r}_{k i}|$ between particles $i$ and $k$, and on the search radius $h_i$ of particle $i$. For typical choices of kernel functions, $W$ takes its maximum at $r_{k i} = 0$ and decreases monotonically as $r_{k i}$ increases. For case b), $\mathbf{A}_{k i}$ is the vector-oriented area of a face constructed between particles $i$ and $k$, in such a way that the entirety of these faces around particle $i$ forms a convex hull. The vector $\hat{\mathbf{r}}_{k i} = \mathbf{r}_{k i} / r_{k i}$ is the unit vector pointing from particle $k$ to particle $i$.
\par We first consider case a): for this simple choice, the bidirectional search boils down to a standard search within radius $h_i$ since the compact support of kernel function $W$ is contained within a radius of $h_i$.
The sum in the denominator on the right hand side of equation \eqref{eq:energy_i_j} can be interpreted as the gas density evaluated at DM particle $j$, and the energy assigned to each gas particle is proportional to its contribution to it. Therefore, the larger the mass of a gas particle and the closer it is to the injecting DM particle, the higher the amount of energy it receives. A variant would be to drop the explicit dependence on the gas particle masses and to set the weights to $w_k = W(r_{k i}, h_i)$. 
However, as most simulation codes use similar (or even identical) gas particle masses, there is typically little change.
\par Whereas in case a), the direction of energy injection around a DM particle depends on the mass distribution of neighbouring gas particles (see Figure \ref{fig:Injection_schemes}), it is often desirable to inject energy in a uniform way without giving preference to any specific direction -- a property dubbed as ``statistical isotropy''. For this purpose, we follow the idea in \citet{HopkinsSupernovae} to use solid angle based weights in case b), which are chosen such that the energy assigned to each particle is proportional to the solid angle that it subtends on the sky as viewed from particle $i$. For more details, in particular on the construction of the convex hull, and an illustrative sketch, we refer the reader to \citet{HopkinsSupernovae}. In contrast to the reference, it is not necessary to define vector weights in our case because the transferred quantity (energy) is a scalar, although this method naturally lends itself to momentum injections as well. In Section \ref{sec:results}, we will investigate how the choice of the weights affects the simulation results.
\par Additionally, more complex weights could be chosen for modelling the gradual deposition of energy along each line of sight. To this end, the sky as seen from a DM particle could be subdivided into multiple viewing cones along each of which a part of the energy would be distributed by setting the weights for particles within the viewing cone. More elaborate energy injection mechanisms will be addressed in future work.
\subsection{Individual time stepping}
\begin{figure}
  \centering
  \noindent
  \resizebox{\columnwidth}{!}{
  \includegraphics{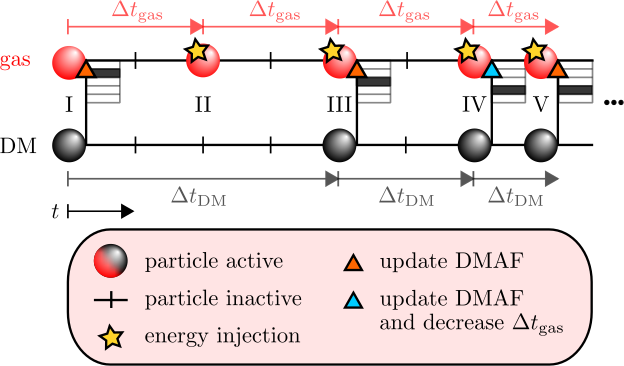}
  }%
  \caption[]{%
Time line of a DM and a neighbouring gas particle: \\ %
I: DM assigns a DMAF energy rate to gas according to equation \protect{\eqref{eq:energy_i_j}}. \\ %
II: End of gas time step, gas particle adds the energy set at I. \\ %
III: End of gas and DM time step, gas adds again the energy set at I. DMAF energy rate is updated and assigned to gas particle. DM reduces its time step, which is now the same as gas time step. \\ %
IV: End of gas and DM time step, gas adds the energy set at III. DM wants to change to a smaller time step than gas. DM informs gas, gas decreases its time step as well. DMAF energy rate is updated and assigned to gas particle. \\ %
V: End of gas and DM time step, gas adds the energy set at IV. DMAF energy rate is updated and assigned to gas particle. \\%
}
\label{fig:Time_stepping}
\end{figure}
Modern cosmological simulation codes such as \textsc{Gizmo} feature an adaptive, individual time stepping scheme which leads to an appreciable computational speed-up as compared to fixed time step schemes, while resolving active regions at high accuracy in time. In \textsc{Gizmo}, the time domain is subdivided in a division of powers of two and after each time step, every particle determines its subsequent time step depending on various time step criteria. These include a criterion for gravitational acceleration proposed by \citet{Power2003} and a Courant--Friedrichs--Lewy (CFL) criterion \citep{CFL1928} for hydrodynamics, together with the improvements in \citet{Saitoh2009, Durier2012, Hopkins2013}. We refer the reader to \citet{Springel2010, GIZMO} for a detailed description of the time stepping scheme in the \textsc{Gadget} codes and in \textsc{Gizmo}. 
\par For the implementation of DMAF, one must hence pay attention to the cases where a receiving gas particle has a smaller or larger time step than the injecting DM particle, i.e. $\dtg \neq \dtdm$. The case $\dtg > \dtdm$ is rather unlikely since the gas particles are subject to additional hydrodynamic time step restrictions, whereas DM is typically subject only to the gravitational time step limiter. For this reason, and since the DMAF energy absorbed by a gas particle may significantly change its dynamics, we opt for the following strategy: if a DM particle is assigned a smaller time step than a receiving gas particle, the gas particle reduces its time step such that $\dtg \leq \dtdm$. However, it can occur that a DM particle encounters a receiving gas particle with a larger time step that is currently inactive; this case is treated in Appendix \ref{sec:inactive_gas}.
\par We devised a first-order in time scheme for the DMAF energy injection that is able to deal with differing time steps for injecting DM and receiving gas particles. An illustration of the scheme is presented in Figure \ref{fig:Time_stepping}. At the beginning of each time step, just after the time steps for the coming step have been found, each active DM particle computes the local DM density, finds the receiving gas particles, and evaluates equation \eqref{eq:energy_i_j} (for the first time at time I in Figure \ref{fig:Time_stepping}). In case of a gas particle with $\dtg > \dtdm$, the respective gas particle will update its new time step to equal $\dtdm$. In the regular case where $\dtg \leq \dtdm$, the receiving gas particle stores the energy rate in a bin corresponding to the time step of the injecting DM particle, illustrated by the black filling of the respective array element.
\par The motivation for the time-binned energy storage is as follows: if a gas particle had a single variable to store the DMAF energy it receives, each DM particle would need to store a list of receiving gas particles in order to let them know when it gets active again. In the example in Figure \ref{fig:Time_stepping}, suppose that the gas particle receives energy from \emph{another} DM particle at time I, which has the same time step as the gas particle. Then, this other DM particle updates its energy injection into the gas at time II (which effectively means that the old energy rate generated by this DM particle should be deleted and be replaced by a new energy rate), whereas the DM particle depicted in the figure is not active at time II and the energy injection from this particle calculated at time I should be sustained. This requires that the other DM particle inform the gas particle about the new energy rate (which might be zero if the gas particle is no longer a neighbour of the DM particle). Since the gas particle has moved during the time step and may now reside within the computational domain of another process, it would be computationally expensive and intricate for the DM particles to keep track of their receivers. 
\par For this reason, we store the DMAF energy rates for each gas particle in an array where each element corresponds to energy from a certain DM time step \footnote{Although codes such as \textsc{Gizmo} and \textsc{Gadget} internally work in terms of specific energy, it is important to store the energy rate $dE/dt$ since the particle mass may change during the time step, e.g. by other feedback processes, or intrinsically due to the hydrodynamic method in case of the meshless finite volume method.}. Thanks to the elegant time discretisation in powers of two, it is then easy to zero out the energy rate generated by particles that will become active. 
\par This is schematically shown in Figure \ref{fig:Time_stepping}: at the end of the first time step of the gas particle, it adds the energy $\frac{dE_{i \to j}}{dt} \dtg$. Since the energy rate from the DM particle has been stored in a bin corresponding to the DM particle's larger time step, it is not zeroed out yet. At time III, the gas particle adds again the energy $\frac{dE_{i \to j}}{dt} \dtg$ set at time I as the DM particle did not update its injection rate at time II. Now, the DM particle is active again and, in this example, reduces its time step. The gas particle zeroes out the energy rate from the DM particle (note that this is without the DM needing to inform the gas particle, but rather because the gas particle ``knows'' that the DM particle is active now since the energy rate is stored in the respective bin). The gas particle is still a neighbour of the DM particle and stores the energy rate from the DM particle in the bin corresponding to the new DM time step $\dtdm$. Note that if the gas particle were not a neighbour of the DM particle anymore, it would simply delete the energy rate from the DM particle set at time I and not set a new contribution to the energy rate from this DM particle. At time IV, the gas particle adds the energy $\frac{dE_{i \to j}}{dt} \dtg$ and zeroes out the energy rate from the DM particle. Now, the DM particle further reduces its time step which might now be smaller than the designated time step of the gas particle. As the DM particle sets the new energy rate for the gas particle, it informs the gas particle that it needs to decrease its coming time step such that it equals the DM time step. At time V, the gas particle adds the energy, zeroes out the energy rate, the DM particle sets the new energy rate, and so on. 
\par In case of $\Delta t_\text{DM} \gg \Delta t_\text{gas}$, gas particles may continue receiving energy from a DM particle although the gas particles have already moved a large distance and are no longer in the vicinity of the injecting DM particle. In order to prevent this, we implemented an option to limit the DM time step to $\Delta t_\text{DM} \leq c_\Delta \Delta t_\text{gas}$ for all receiving gas particles, in the spirit of \citet{Saitoh2009} who recommend $c_\Delta = 4$ for neighbouring gas particles. This enforces that DM particles update their gas neighbours more frequently, leading to a more localised energy injection. 
\par In cosmological simulations, proper treatment of the cosmological expansion is required, which we discuss in Appendix \ref{sec:cosmological_expansion}.

\subsection{Sedov--Taylor time step limiter}
\label{subsec:Sedov_Taylor}
Assume a gas particle moves from a low-density DM region to a high-density region. Since the DMAF power in the low-density region was small, it might still have a large time step, at the end of which it adds a big amount of energy from DMAF. Since the energy is directly injected as internal energy, the time step criteria that depend on dynamic quantities such as the particle acceleration remain unaffected and the gas particle will carry out another large time step, during which parts of the energy are converted into kinetic energy. This means that two (possibly) large time steps may pass from the moment the gas saves the new DMAF energy rate until it reduces its time step.
\par In order to react faster to the increase in DMAF energy, we harness the fact that the total DMAF energy rate for each gas particle is already known at the beginning of the time step. Define the set of donor particles as $\hat{\mathcal{N}}_\text{DM}(j) = \{i \in \text{DM}: j \in \mathcal{N}_\text{gas}(i) \}$. Let $\mathcal{P}_j = \sum_{i \in \hat{\mathcal{N}}_\text{DM}(j)} \frac{dE_{i \to j}}{dt}$ be the total DMAF power that particle $j$ receives. If $\mathcal{P}_j$ is sufficient to form a strong shock, the shock propagation will only depend on $\mathcal{P}_j$ and the surrounding gas density $\rho_g$ as the surrounding gas pressure is negligible. Using dimensional analysis, the shock radius $R_{s,j}$ for this generalised Sedov--Taylor self-similar shock can be found to propagate as
\begin{equation}
R_{s,j} = \beta \left(\frac{\mathcal{P}_j}{\rho_{g,j}}\right)^{\nicefrac{1}{5}} t^{\nicefrac{3}{5}},
\end{equation}
approximating the surrounding gas density in vicinity to particle $j$ by the constant value $\rho_{g,j}$, and assuming constant DMAF power during the time step. The dimensionless coefficient $\beta$, which depends on the polytropic index of the gas $\gamma$, can be determined numerically after a lengthy calculation to be of magnitude $\sim 1$ (see \citealt{Dokuchaev2003} for exact values), for which reason we neglect it in what follows.
\par In the spirit of a CFL condition, we restrict the time step to be small enough such that a strong shock in a (locally) sufficiently homogeneous environment is confined to the smoothing length $h_j$ of particle $j$ until the end of the time step, that is
\begin{equation}
R_{s,j} \leq c_\text{CFL} h_j,
\end{equation}
where $c_\text{CFL}$ is the CFL number. This yields an additional time step limiter of
\begin{equation}
\Delta t_j \leq \left(c_\text{CFL} h_j\right)^{\nicefrac{5}{3}} \left(\frac{\rho_{g,j}}{\mathcal{P}_j}\right)^{\nicefrac{1}{3}},
\end{equation}
which we impose at the beginning of each gas time step just after the DMAF power has been calculated which might lead to an updated smaller time step. 
\par In none of our numerical tests for a homogeneous, isotropic universe and a halo with a gaseous fraction, this Sedov--Taylor time step limiter was the dominant time step criterion, while for our academic example with extremely high annihilation rates, it turned out to be necessary in order to prevent big energy errors, which we will discuss in Subsection \ref{subsec:step_example}.

\subsection{Function flow}
\begin{figure}
  \centering
  \noindent
  \resizebox{\columnwidth}{!}{
  \includegraphics{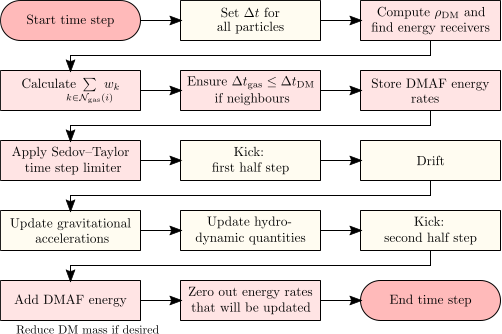}
  }%
  \caption{Flowchart for one time step of the DMAF implementation. Rectangles tinted in light pink stand for steps that are required due to the DMAF, whereas light yellow rectangles indicate steps that are always required. Recall that the kick evolves the system in momentum space while the drift operation updates the positions.}%
\label{fig:Flowchart}
\end{figure}

Figure \ref{fig:Flowchart} provides a schematic overview of the DMAF method. Steps unrelated to the DMAF are sketched very roughly only to provide an overview. Note in particular the logic for determining the gas time step: first, $\Delta t$ is set for \emph{all} particles, irrespective of DMAF. Then, the energy rates at the DM particles are computed and energy receivers are determined. If a DM particle has a smaller time step then a receiving gas particle, the latter reduces its time step. The energy rates are stored for each gas particle. Afterwards, gas particles check if the Sedov--Taylor time step condition is satisfied, otherwise, they reduce their time step. Then, the system evolves to the end of the time step by applying two kick half steps and a drift; between the kick half steps, gravitational forces and hydrodynamic quantities are updated. At the end of the time step, gas particles add the energy from DMAF and zero out the energy rates from DM particles that will subsequently update their energy rates.

\section{Donor-based vs. receiver-based approach}
\label{sec:comparison}
Whereas in our implementation, the energy deposition is expressed by equation \eqref{eq:energy_i_j}, another way of incorporating DMAF energy in cosmological simulations is proposed in \citet{Iwanus2017}. In that work, the energy generated by DMAF is determined at each gas particle, for which reason we will refer to this method as receiver-based approach in what follows. This section is dedicated to a brief comparison between the receiver-based approach and the donor-based approach proposed herein, highlighting the strengths and weaknesses of either of them. 
\par In the receiver-based approach, equation \eqref{eq:energy_rate} is reformulated as
\begin{equation}
\frac{1}{M_g} \frac{dE}{dt} = \frac{\langle\sigma v\rangle}{m_\chi} \frac{\rho_\chi^2}{\rho_g} c^2,
\end{equation}
where $M_g$ denotes the gas mass contained in the fixed volume $V$ and $\rho_g$ stays for the gas density. From this, the energy rate at a gas particle $j$ produced by DMAF can be computed as
\begin{equation}
\frac{1}{M_j} \frac{dE_{j\leftarrow \chi}}{dt} = \frac{\langle\sigma v\rangle}{m_\chi} \frac{\rho_{\chi,j}^2}{\rho_{g, j}} c^2.
\label{eq:energy_i_j_receiver_based}
\end{equation}
In equation \eqref{eq:energy_i_j_receiver_based}, $E_{j\leftarrow \chi}$ is the total energy received by particle $j$, $\rho_{\chi, j}$ and $\rho_{g, j}$ are the DM density and gas density evaluated at particle $j$, respectively, and $M_j$ is the mass of particle $j$.
\par The receiver-based approach underlies the assumptions that the DM particles can be used as tracers to reconstruct a well-defined DM density field and that the energy injection is highly localised.
\subsection{Energy injection}
First, let us compare the total energy rates created by DMAF in a constant volume for both methods. For the ease of notation, we assume that the kernel function $W = W(r, h)$ is independent of the particle type under consideration. A short calculation gives
\begin{equation}
\label{eq:energy_rate_ratio}
\frac{\left(dE/dt\right)_\text{donor}}{\left(dE/dt\right)_\text{receiver}} = \frac{\sum_{i \in \text{DM}} M_i \left(\sum_{k \in \mathcal{N}_\text{DM}(i)} M_k W(r_{ki}, h_i) \right)}{\sum_{j \in \text{gas}} \left( M_j \frac{\left(\sum_{k \in \mathcal{N}_\text{DM}(j)} M_k W(r_{kj}, h_j)\right)^2}{\sum_{l \in \mathcal{N}_\text{gas}(j)} M_l W(r_{lj}, h_j)}\right)},
\end{equation}
where $\mathcal{N}_T(i)$ denotes the set of neighbours of type $T \in \{\text{gas}, \text{DM}\}$ of particle $i$ and where we used the SPH density estimate $\rho_i = \sum_{k \in \mathcal{N}(i)} M_k W(r_{ki}, h_i)$. The total energy rate in the donor-based scheme thus does not depend on either gas properties nor on the choice of weights in equation \eqref{eq:energy_i_j}, whereas the evaluation of DMAF at the gas particles in the receiver-based case implies that changes in the gas distribution yield a change in the total DMAF energy rate. The enumerator, originating from the donor-based method, only depends on distances between DM particles in the evaluation of the local DM density, while the denominator shows that for the receiver-based method, distances between gas and DM particles as well as between gas particles determine the energy rate.
\par The independence of the total DMAF energy rate from the gas properties is arguably an advantage of the donor-based approach. Since for cold WIMP-like particles, the impact of the energy generated by DM annihilation outweighs the effects of the DM mass loss itself by far \citep{Iwanus2017}, it is a reasonable approximation to keep the DM particle masses constant throughout the simulation. However, if the mass loss due to annihilation is to be taken into account explicitly, e.g. for light dark matter candidates, the donor-based approach allows for reducing the DM mass with $e = M c^2$ being satisfied to machine accuracy by simply subtracting the DM mass corresponding to the DMAF energy at the end of the time step. In contrast, the DM mass loss in the receiver-based implementation is decoupled from the energy absorption calculated at the gas particles, which introduces a small mass error. 
\par Additionally, if a more detailed prescription of the energy deposition is required where one wants to account for a certain mean free path of the annihilation products, the donor-based approach provides the flexibility to adjust the weights to mimic the desired physics while the receiver-based approach has no free parameters. This is crucial since the particular annihilation channel may significantly impact the total energy absorbed by the surrounding gas (see e.g. \citealt{Schon2017}).
\par On the other hand, the receiver-based approach intrinsically localises the energy deposition at the gas: if a DM particle resides in a region without gas particles, no energy from this DM particle will be absorbed since no receivers are nearby. In particular, energy production from DMAF in extremely gas-poor haloes is artificially suppressed with the receiver-based method, while the donor-based method sustains the energy injection into gas particles even after the DMAF has driven them out of their host halo if not enough gas particles are left within the halo that could absorb the energy. This means that with the donor-based method, theoretically, distant gas particles might instantaneously receive large amounts of energy because the energy created at the DM particles needs to go \emph{somewhere}. However, in cosmological simulations where $N_\text{gas} \approx N_\text{DM}$, the distance between each DM particle and its nearest gas neighbours is typically small and unphysical energy transport is unlikely to occur. Additionally, relativistic annihilation products travel distances of $\sim 100 \ \text{kpc}$ within a typical time step of a few hundred thousand years, for which reason the physical energy transport horizon is well above the spacing of gas particles in dense regions and the assumption of instantaneous energy injection into the nearest gas neighbours seems justified.
\par In both methods, incorporating arbitrary absorption fractions $f$ and boost factors $B$ can be done without any difficulty. 

\subsection{Computational cost}
\begin{table}
\begin{center}
\begin{tabular}{@{}llcccc@{}}
\toprule
   & Neighbour search                         & \multicolumn{1}{l}{donor} & \multicolumn{1}{l}{receiver} & search at & search for \\ \midrule
1. & Calculate $\rho_g$                  & X                               & X                                  & gas       & gas        \\
2. & Calculate $\rho_{\chi, i}$       & X                               &                                    & DM        & DM         \\
3. & Calculate $\rho_{\chi, j}$ &                                 & X                                  & gas       & DM         \\
4. & Find energy receivers        & X                               &                                    & DM        & gas        \\ \bottomrule
\end{tabular}
\caption{Different neighbour searches are required for the donor-based and for the receiver-based approaches, respectively. The gas density is always calculated irrespective of DMAF. The two methods differ in the location where the DM density is evaluated. Finding energy receiving gas neighbours is only necessary for the donor-based approach.}
\label{table:loops}
\end{center}
\end{table}
The implementation of DMAF necessitates additional neighbour searches for different particle types, which are listed in Table \ref{table:loops} for the donor-based and receiver-based approach. The first neighbour search for $\rho_g$ is always needed regardless of the DM annihilation. Thanks to the similarity of the neighbour searches, large parts of the code can be recycled from the gas density calculation. The extra neighbour searches can be expected to constitute the major part of additional cost for the DMAF implementation. Note, however, that as reported in \citet{GIZMO}, the number of iterations needed until the search radius has converged is very small for the gas density, and we observe the same behaviour for the other loops as well. This is due to the efficient iteration scheme by \citet{SpringelHernquist2002} that has been further improved in \textsc{Gizmo} accounting for findings in \citet{Cullen2010, Hopkins2013, FIRE1}. 
\par The donor-based approach as given by equation \eqref{eq:energy_i_j} is based upon the evaluation of $\rho_\chi$ at each DM particle, which requires a search of the nearest DM neighbours. Moreover, a neighbour search must be conducted for each DM particle in order to find the receiving gas particles. Both neighbour searches are conducted in analogy to the calculation of the gas density at each gas particle: since the desired number of neighbours is fixed (and not the search radius), a variable search radius $h_i$ is assigned to each DM particle $i$ which is adjusted until the effective neighbour number, calculated via $N_{\text{ngb, eff}} = (4\pi/3) \, h_i^3 \sum_k W(r_{k i}, h_i)$, has reached the desired value up to a certain tolerance. Separate search radii $h_{i, g}$ and $h_{i, \chi}$ for gas neighbours and DM neighbours, respectively, are needed, and two neighbour searches must be carried out. In total, there are thus three neighbour search loops. The loop over the energy receiving gas particles is in fact called twice: in the first call, the weight for each gas particle is calculated, then, the normalisation factor in the denominator of equation \eqref{eq:energy_i_j} is computed, and in the second call, the normalised DMAF energy rates are stored. However, note that gas particles often possess a smaller time step than the DM particles since the time step for gas particles is subject to a range of limiters depending on the baryonic physics (see \citealt{GIZMO}), for which reason the additional loops may be evaluated less frequently than the gas density loop in many situations. 
\par The overhead from updating the gas time step such that $\Delta t_\text{gas} \leq \Delta t_\text{DM}$ is satisfied for neighbouring particles and from applying the Sedov--Taylor time step limiter is small.
\par For the receiver-based approach, in contrast, all properties in equation \eqref{eq:energy_i_j_receiver_based} are being determined by default, except for the DM density at gas particles. Hence, one additional neighbour search is required. In case of a small ratio between gas time steps and DM time steps, however, the less frequent evaluation of the two loops in the donor-based approach may outweigh the receiver-based approach in terms of computational speed. 
\par In the first example in Section \ref{sec:results}, the donor-based method with choice of weights a) was the fastest, followed by the receiver-based method and the donor-based method with choice of weights b). For the realistic examples, the receiver-based method outperformed the donor-based method in terms of computational speed. For the cosmological simulation for example, run on 256 CPUs, the wall time was $7965$, $13547$, $16700$, $6387 \ \text{s}$ for $\Lambda \text{CDM}$, donor-based a), donor-based b), and receiver-based -- thus, the receiver-based method ran even faster than the fiducial simulation without DMAF.

\subsection{Time stepping}
\par In the receiver-based approach, each gas particle updates the DMAF energy rates together with the hydrodynamic quantities between the two kick half steps (see Figure \ref{fig:Flowchart}) and adds the DMAF contribution to the total energy rate. Thus, the DMAF energy evolves with the second-order in time leapfrog method as a component of the total energy budget.
\par For the donor-based method, we use a simple first-order accurate time stepping scheme. The motivation for this is the following: since the DMAF energy rates are computed at the DM particles, the update of the DMAF energy rate needs to happen at a logical moment during each DM time step. One could update the DMAF energy rate between the two half kicks of each DM particle, however, the length of the DM half kicks generally differs from the one of the receiving gas particles due to different time steps. Therefore, this would lead to an update of the DMAF energy rate in an unnatural moment \emph{for the gas} (e.g. in Figure \ref{fig:Time_stepping} before the gas particle performs its second half kick from the midpoint between II and III to III, the DM particle, which is about to do its second half kick from II to III, would update the DMAF energy rate, which would only affect one out of four half kicks of the gas between times I and III).
\par For this reason, we opt for the aforementioned method consisting of updating the DMAF energy rates at the beginning of each DM time step (in case the DM particle is active) and adding the energy at the end of each gas time step. In view of the large uncertainties in the DM annihilation mechanism, employing a simple first-order in time scheme for the DMAF energy seems justifiable.

\section{Results}
\label{sec:results}
In this section, we present test cases for our numerical method. For all tests, we use the meshless finite mass method (MFMM), which is the default method in \textsc{Gizmo}. We take a standard cubic spline kernel. In our tests, we neglect modelling the small DM mass loss due to annihilation.
\par The first example deals with the injection of DMAF energy into a contact discontinuity. Then, we turn towards the more realistic case of an isolated halo. Finally, we consider the effects of DMAF in a cosmological simulation of non-linear structure formation. 

\subsection{Injection into gas with a density jump}
\label{subsec:step_example}
Our first numerical example examines the differences arising from the donor-based approach and the receiver-based approach. Additionally, the two different choices of weights considered herein are compared. The scenario for the simulation is the following: a few massive DM particles are located adjacent to a discontinuous transition between a high-density gas region on the right and a low-density gas density region on the left (density ratio $10^4 \colon 1$). The geometry is two-dimensional and the box is large enough that the shock ensuing the DMAF energy deposition into the gas remains within the simulation domain until the end time of the simulation $T = 978.5 \ \text{Myr}$. 
\par Initially, the gas is distributed over a regular grid consisting of $256 \times 256$ points. All gas particles located at $x < 0$ possess a mass of $M = 1.5 \times 10^7 \ \text{M}_\odot$, whereas all gas particles in the right hand side of the domain ($x > 0$) have a mass of $M = 1.5 \times 10^{11} \ \text{M}_\odot$. This results in a density jump of four orders of magnitude across $x = 0$ since the particles are equally spaced over the simulation domain ($\sim 10^6 \ \text{M}_\odot \ \text{kpc}^{-3}$ and $\sim 10^{10} \ \text{M}_\odot \ \text{kpc}^{-3}$ on the left hand side and right hand side, respectively).
\par
\begin{figure}
  \centering
  \noindent
  \resizebox{\columnwidth}{!}{
   \includegraphics{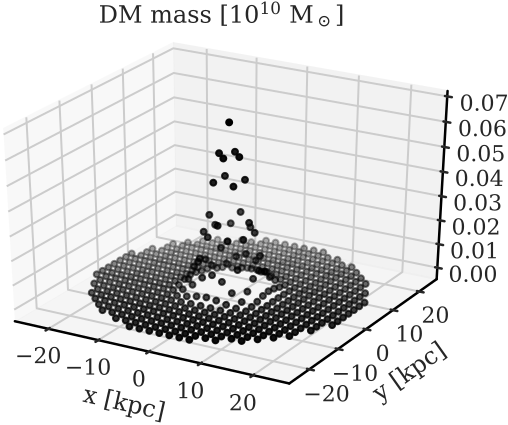}
  }%
  \caption{Static mass distribution of the DM particles, following the PDF of a normal distribution with mean $(0, 0) \ \text{kpc}$ and standard deviation $3 \ \text{kpc}$. Shown are particles with mass larger than $5 \times 10^{-6} \ \text{M}_\odot$.}
\label{fig:Step_example_ICs}
\end{figure}

 We populate the simulation domain with $513 \times 513$ DM particles whose masses follow the probability density function (PDF) of a two-dimensional normal distribution with mean $(0, 0) \ \text{kpc}$ and standard deviation $3 \ \text{kpc}$, scaled such that the total DM mass in the system equals $10^{10} \ \text{M}_\odot$ (see Figure \ref{fig:Step_example_ICs}). The number of DM particles is chosen such that the particle with maximal mass resides in the origin of the domain, exactly in the middle between the nearest gas particles in the left and right domain half. The larger number of DM particles as compared to the gas is taken to have sufficiently many particles in the small region where the high DM density leads to a substantial DMAF energy generation. A minimum particle mass of $10^{-6} \ \text{M}_\odot$ is enforced to ensure numerical stability. Since the DM mass distribution declines rapidly as the distance to the jump increases, the DMAF energy deposition will be dominated by a small number of DM particles around the origin.
\par Depending on the injection mechanism, we expect differing energy fractions and propagation speeds of the emanating shock wave in each domain half. For the receiver-based method, which assumes a well-defined DM density field, this problem is numerically challenging in view of the steep DM density gradients and the crude approximation of the underlying smooth DM density field by the DM particles. Self-gravity is switched off for this test case; moreover, the gas is initially cold such that the system starts in an equilibrium state. We take the gas to be polytropic with $\gamma = 5/3$. For the DM candidate, we choose a thermal relic cross-section of $\langle \sigma v \rangle = 3 \times 10^{-26} \ \text{cm}^3 \ \text{s}^{-1}$ and a particle mass of $m_\chi = 100 \ \text{keV} \ c^{-2}$. The number of desired neighbours is $N_\text{ngb} = 16$ for all neighbour searches.
\begin{figure*}
  \centering
  \noindent
  \resizebox{\textwidth}{!}{
   \includegraphics{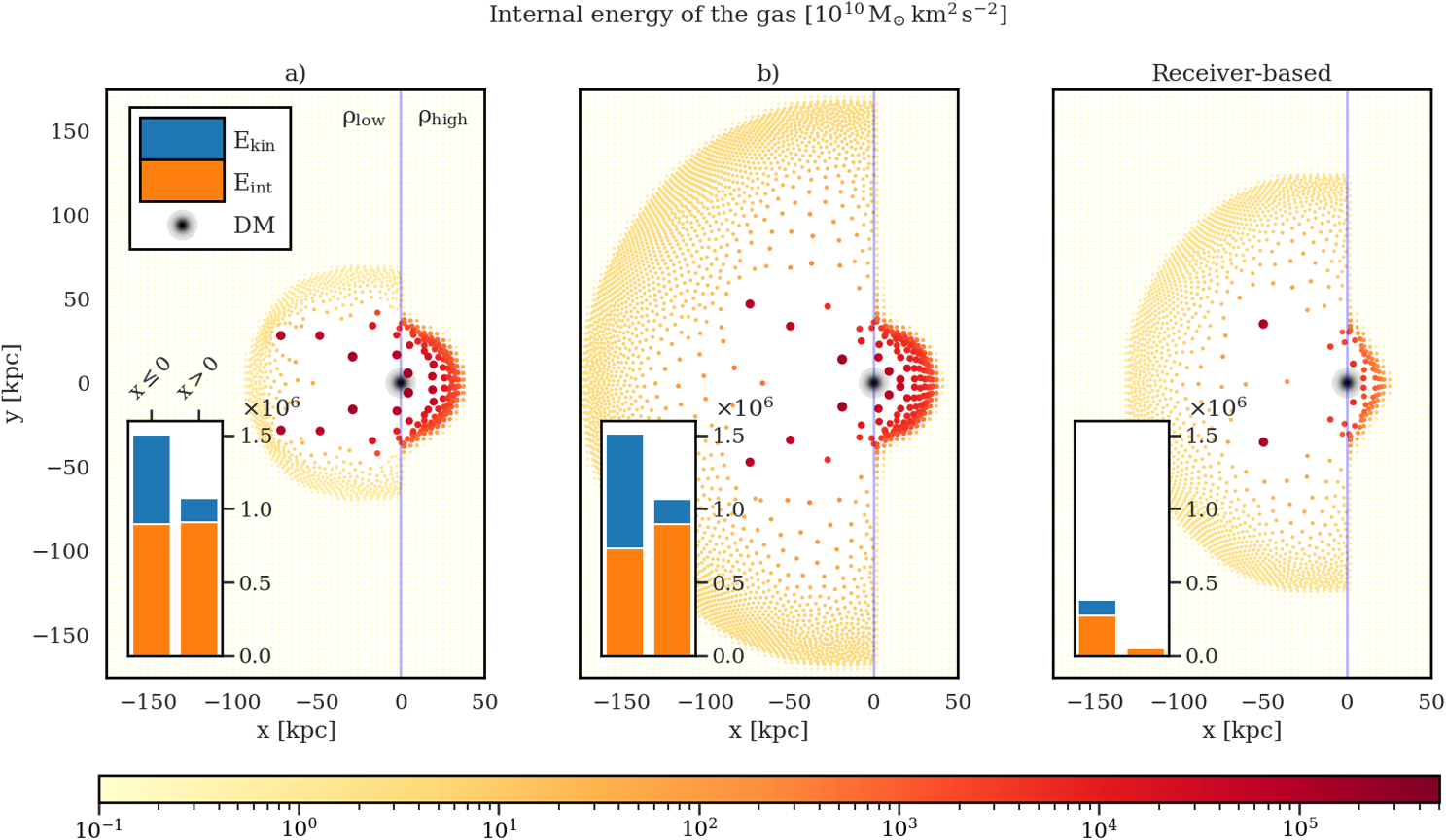}
  }%
  \caption{Internal energy of the gas for the donor-based method with choice of weights a) (\emph{left}), b) (\emph{centre}), and receiver-based method (\emph{right}), at the end time of the simulation. Each filled circle stands for one particle and the size and colour represent the internal energy. The spherical black area encompasses the $3\sigma$-region of the DM density field for which the DM particles act as a tracer. The blue line marks the density jump. The kinetic and internal energies in the left and right half of the domain are depicted in the inset bar plots. In the solid angle weighted variant b), a bigger fraction of the energy in the left low-density subdomain has been converted to kinetic energy, and the shock wave has propagated further than in the mass weighted variant a). The total energy injected by the receiver-based approach is much smaller.}
\label{fig:Step_example_result}
\end{figure*}
\par Figure \ref{fig:Step_example_result} shows the internal energy of the gas at final time. Each dot represents a gas particle, where particle size and colour scale with the internal energy of the particle. The black spherical region around the origin marks the $3\sigma$-region of the PDF which the DM particles discretise. Because of the lower density in the left subdomain, the shock speed is faster than in the right subdomain, which can be seen for all methods. In the case of mass weighted injection (case a)), the shock front has propagated the least into the left, low-density subdomain. Since the gas density in the right subdomain is much higher than the one in the left subdomain, the DM particles preferentially deposit DMAF energy into the right subdomain. Tracing back the movement of the high-energy particles in the left subdomain, one finds that they originate from the right domain and migrate to the left half after having absorbed a large amount of energy. 
\par In the solid angle weighted case b), the shock front in the left subdomain has advanced much further. Due to the statistical isotropy of the energy injection, the energy injection is not biased towards the right subdomain. 
\par For the receiver-based approach, the shock front in the left subdomain lies between cases a) and b). It is salient that the shock wave in the right subdomain is much less pronounced than for the the donor-based approach; the shock front has travelled a shorter distance and the shock carries a much lower energy. Also, the total number of high-energy particles is much smaller.
\par The inset plots show the energy in the left and right half of the domain, broken down into internal energy and kinetic energy.  For all methods considered, the energy in the left subdomain at final time is greater than the one in the right one due to convective energy transport. For the two flavours of the donor-based method, the total energy in the system should be identical since the methods only differ in \emph{how} the energy is deposited into the gas, not in the calculation of the DMAF energy rate. Indeed, the difference in total energy after $978.5 \ \text{Myr}$ between case a) and b) amounts to $< 10^{-4}$ per cent. In contrast, the total energy for the receiver-based method is only $17.3$ per cent of the one for the donor-based method. This is caused by the reconstruction of the DM density at the gas particles, which leads to an underestimation of the DMAF energy rate: while equation \eqref{eq:energy_i_j} for the donor-based method depends linearly on $\rho_\chi$, equation \eqref{eq:energy_i_j_receiver_based} has a quadratic dependence on $\rho_\chi$ and therefore on an average over several DM particles, which suppresses sharp mass peaks of individual DM particles and hence lowers the generated DMAF energy.
\par Interestingly, the shock wave has propagated further than for the donor-based method in case a), despite the much lower total energy. This is due to the fact that in the receiver-based method, gas particles in the left subdomain absorb DMAF energy throughout the simulation, whereas for the mass weighted injection, the largest part of the energy in the left domain half comes from particles that have propagated from the right domain half to the left, and the fraction of directly absorbed energy in the left domain half is small. During the first $196 \ \text{Myr}$, the energy in the left domain half is larger with the receiver-based method than with the mass weighted donor-based method.
\begin{table}
\begin{center}
\emph{With} time step limiter: \\
\begin{tabular}{@{}llll@{}}
\toprule
                              & a)      & b)       & Receiver-based \\ \midrule
$E_\text{tot}(t_1)$              & 257840  & 257650   &     173910            \\
1. -- 3. $\Delta t_\text{min}$   & 0.11936 & 0.014920 &     0.014920          \\ \bottomrule
\end{tabular}
\vspace{0.5cm}
\\
\emph{Without} time step limiter: \\
\begin{tabular}{@{}llll@{}}
\toprule
                           & a)      & b)       & Receiver-based \\ \midrule
$E_\text{tot}(t_1)$        & 260700  & 309830   &    233010      \\
1. $\Delta t_\text{min}$   & 15.278  & 15.278   &    15.278      \\
2. $\Delta t_\text{min}$   & 15.278  & 15.278   &    0.029840    \\
3. $\Delta t_\text{min}$   & 0.11936 & 0.029840 &    0.029840    \\ \bottomrule
\end{tabular}
\caption{Total energy in the system at $t_1 = 98 \ \text{Myr}$ and the first three system time steps. Energies are given in $10^{10} \ \text{M}_\odot \ \text{km}^2 \ \text{s}^{-2}$ and times in $\text{Myr}$. With the time step limiter, the total energies for variant a) and b) agree well with each other, while there is a large energy error for variant b) and for the receiver-based method without the time step limiter. The time step imposed by the additional limiter \emph{before} the energy injection is identical to (for variant a)) or half (for variant b) and for the receiver-based method) the one that is set by the CFL condition \emph{after} the energy injection.}
\label{table:dt_limiter}
\end{center}
\end{table}
\par In order to verify the statistical isotropy in case b), we ran the same simulation with the Euler equation deactivated, i.e. with injection of DMAF energy into static particles. The energy difference between the left and right subdomain at final time then amounts to roughly $0.01$ per cent. Also for the receiver-based approach, the energy is deposited isotropically in this case and the energy difference at final time between the left and right subdomain is less than $0.1$ per cent if the particles are kept fixed. This is because the gas particles are distributed symmetrically around the DM particles and the energy rate in equation \eqref{eq:energy_i_j_receiver_based} is invariant under a scaling of the gas particle mass $M_j$ while keeping the volume occupied by the particle constant. However, this will in general not be the case when the gas is heterogeneously distributed around the DM particles since each gas particle evaluates the DM density at its own location. Finally, for the mass weighted variant of the donor-based method, the energy ratio between the two domain halves at final time is $E_{\text{right}} / E_{\text{left}} = 520$, heavily biased towards the right, high-density domain half.\footnote{We checked that in the case of a single massive DM particle of mass $10^{10} \ \text{M}_\odot$ located at $(0, 0) \ \text{kpc}$ and for static gas particles, the energy ratio for the mass weighted variant differs from the density ratio only by $|E_{\text{left}} / E_{\text{right}} - 10^{-4}| < 10^{-11}$, as expected.}
\par Finally, we discuss the necessity of the Sedov--Taylor time step limiter (see Subsection \ref{subsec:Sedov_Taylor}) to prevent errors at the beginning of the simulation before the deposited internal energy has been converted to kinetic energy. Table \ref{table:dt_limiter} lists the total energy at time $t_1 = 98 \ \text{Myr}$ and the first three system time steps (i.e. smallest time step of any particle), with and without the additional time step limiter. 
\par While the energies for the two choices of weights agree well with each other when the time step limiter is activated, the energy is massively overestimated for solid-angle based weights and for the receiver-based method without the time step limiter. This stems from the fact that these two mechanisms inject large amounts of energy into the left, low-density subdomain, where the energy induces high accelerations which make small time steps necessary. Without the time step limiter, the system performs two huge time steps for the donor-based method (one for the receiver-based method), only limited by the globally set maximum step size, before drastically reducing the time steps. The time step assigned to the particles by the Sedov--Taylor limiter equals or is half the time step enforced by the CFL condition after the energy has affected the dynamics (recall that the time domain is subdivided in powers of two in \textsc{Gizmo}; hence, the minimum time step placed by all applicable limiters is rounded down to the closest power of two times a base interval). Thus, our additional time step limiter may preclude large energy errors while not being overly restrictive.

\subsection{Isolated galaxy}
\label{subsec:halo_example}
\begin{figure*}
  \centering
  \noindent
  \resizebox{\textwidth}{!}{
  \includegraphics{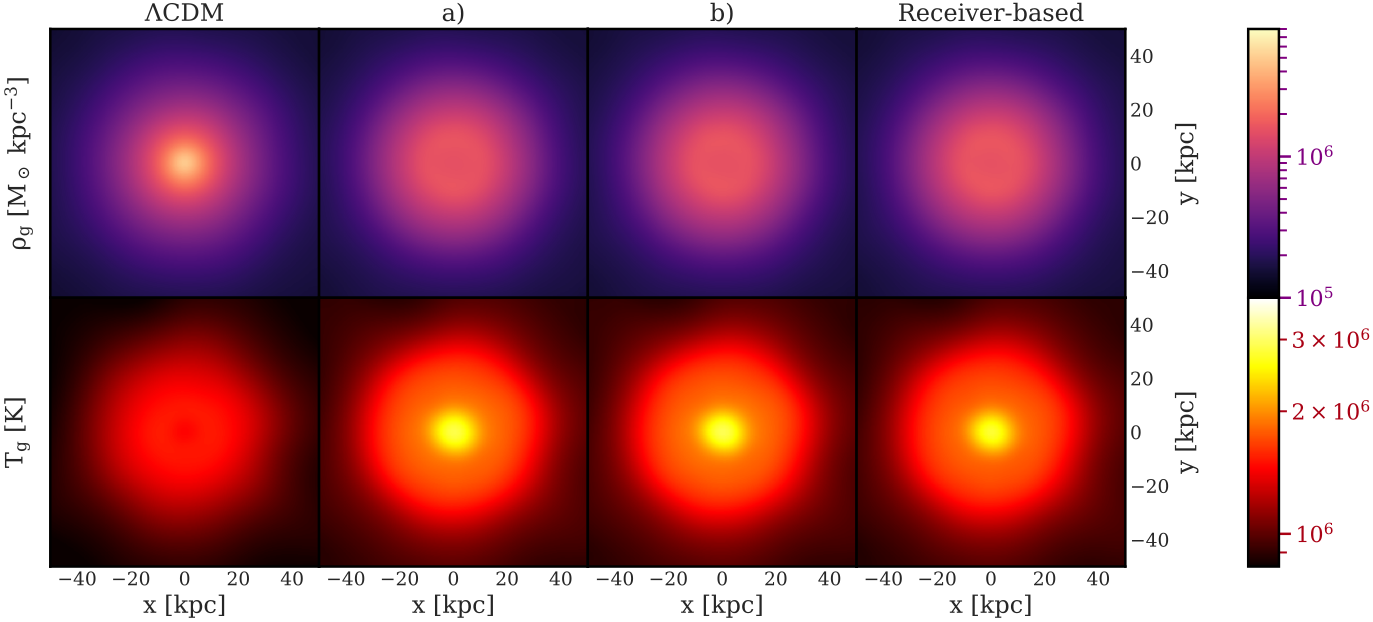}
  }
  \caption{Gas density $\rho_g$ (\emph{top}) and \red{temperature $T_g$} (\emph{bottom}) of the galaxy after $97.8 \ \text{Myr}$, averaged over the $z$-coordinate. The central gas density is depleted due to the DMAF as compared to the fiducial $\Lambda\text{CDM}$ simulation without DMAF. The results for the different methods closely resemble each other.}
\label{fig:Halo_result}
\end{figure*}
\begin{figure*}
  \centering
  \noindent
  \resizebox{\textwidth}{!}{
  \includegraphics{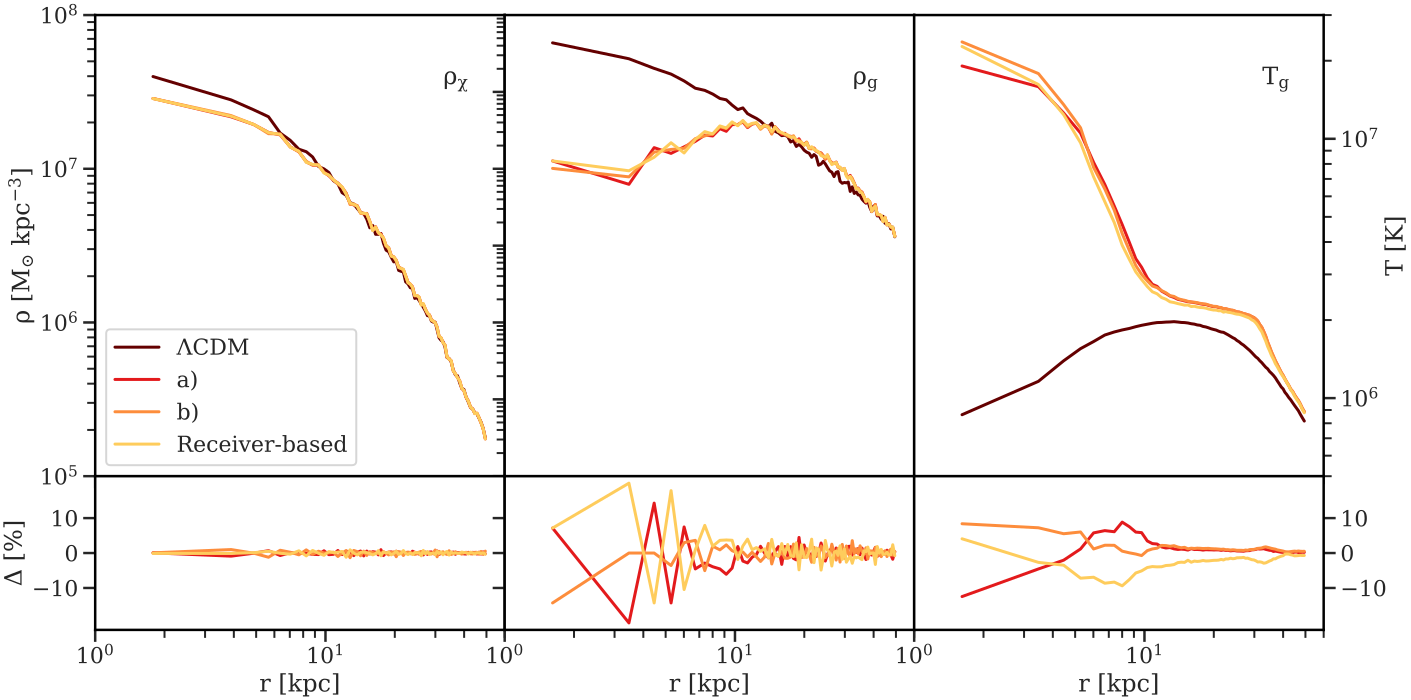}
  }
  \caption{\red{\emph{Isolated halo}: r}adial plot of the DM density $\rho_\chi$ (\emph{left}), gas density $\rho_g$ (\emph{centre}), and \red{gas temperature $T_g$} (\emph{right}) in a logarithmic scale. DMAF reduces the central density of the gas and the DM while increasing the \red{temperature}. The lower panels show the relative difference towards the mean of a), b), and the receiver-based method.}
\label{fig:Halo_radial_plot}
\end{figure*}
Next, we assess our DMAF method in the context of an isolated spherical galaxy, consisting of a DM halo and gas distributed following a Navarro--Frenk--White (NFW) profile \citep{Navarro1997}, with a system mass of $M_{200} = 1.13 \times 10^{12} \ \text{M}_\odot$, concentration parameter $c = 13$, and scale length $r_\text{scale} = 21.7 \ \text{kpc}$. The initial conditions were created using \textsc{Dice} \citep{DICE} and are made up of 80000 DM particles (total mass: $6.04 \times 10^{11} \ \text{M}_\odot$) and 20000 gas particles (total mass: $5.24 \times 10^{11} \ \text{M}_\odot$) that follow the same distribution as the DM and are in thermal equilibrium. We let the initial conditions re-virialise without DMAF for $489 \ \text{Myr}$ to eliminate transients.
\par We simulate the evolution of the galaxy undergoing DM annihilation from a light DM candidate of mass $m_\chi = 100 \ \text{keV} \ c^{-2}$ with thermal relic cross-section of $\langle \sigma v \rangle = 3 \times 10^{-26} \ \text{cm}^3 \ \text{s}^{-1}$ for a simulation time of $T = 97.8 \ \text{Myr}$. The background cosmology is static and thus non-expanding. We take $N_{\text{ngb}} = 40$, and the gravitational softening length is $r_\text{soft} = 4 \ \text{kpc}$. The gravitational softening length plays an important role since it is related to the relative error in the DMAF energy rate due to an unresolved NFW cusp, as derived in \citet{Iwanus2017}.
\par Figure \ref{fig:Halo_result} shows the gas density $\rho_g$ and the \red{temperature $T_g$} of the gas within a radius of $50 \ \text{kpc}$ around the halo centre, averaged over the $z$-coordinate (created with the package \textsc{Pynbody} \citep{pynbody}). In the central region of the halo where the DM density is the highest, the \red{temperature} has risen due to the DMAF energy injection and the gas density has decreased since the heated gas has partly left the halo centre. The depletion of gas in the halo centre also entails a reduction in DM density, as the radial plot in Figure \ref{fig:Halo_radial_plot} shows, albeit not as pronounced as for the gas. The receiver-based and the donor-based method for both choices of weights give very similar results for this example. The relative difference between each of the three simulations with DMAF and their mean amounts to less than $2$ per cent for the DM density, $20$ per cent for the gas density within a radius of $10 \ \text{kpc}$ (less than $5$ per cent for $r > 10 \ \text{kpc}$), and less than $13$ per cent for the \red{temperature}.
\par This example demonstrates that DMAF can considerably alter the structure of galaxies: after less than $100 \ \text{Myr}$, the radial gas density distribution has flattened so much that it is not monotonic anymore but reaches a maximal gas density at a distance of $\sim 10 \ \text{kpc}$ from the galactic centre. For heavier DM candidates, smaller annihilation velocity cross-sections, or lower absorption rates, the imprint of DMAF on the density profiles is smaller but may none the less be detectable, for which reason numerical simulations are a valuable means in investigating the nature of DM. 

\subsection{Cosmological simulation}
In this example, we test our method in the context of a cosmological simulation. The background expansion of the universe requires the conversion of the quantities in equation \eqref{eq:energy_i_j} from comoving to physical coordinates. In order to get a more accurate estimation of the DMAF energy rates, we account for the Hubble flow of inactive DM particles by rescaling the energy rates appropriately. This is discussed in Appendix \ref{sec:cosmological_expansion}. 
\par We simulate a cubic box with side length $50 \ \text{Mpc} \ h^{-1}$ starting from redshift $z = 100$ to $z = 0$. For both DM and gas, we take $256^3$ particles, respectively. 
The boundary conditions are periodic. We select a very light DM candidate here in order to highlight the impact of DMAF. To be specific, we choose $m_\chi = 1 \ \text{MeV} \ c^{-2}$ and a thermal relic cross-section of $\langle \sigma v \rangle = 3 \times 10^{-26} \ \text{cm}^3 \ \text{s}^{-1}$. The background cosmology is taken from \citet{Ade2016} and the parameters are summarised in Table \ref{table:planck2015}.
\begin{table}
\begin{center}
\begin{tabular}{@{}ll@{}}
\toprule
Parameter                                                 & Value  \\ \midrule
$\Omega_m$                                            & 0.3089 \\
$\Omega_b$                                            & 0.0486 \\
$\Omega_\Lambda$                                      & 0.6911 \\
$H_0 \ [\text{km} \ \text{s}^{-1} \ \text{Mpc}^{-1}]$ & 67.74  \\ \bottomrule
\end{tabular}
\caption{Cosmological parameters for the cosmological simulation, taken from \citet{Ade2016}.}
\label{table:planck2015}
\end{center}
\end{table}
For the reconstruction of all hydrodynamic quantities as well as for the number of energy receivers, we choose a desired neighbour number of $N_{\text{ngb}} = 40$. We set the gravitational softening length to $\sim9.77 \ \text{kpc} \ h^{-1}$, which corresponds to 5 per cent of the average inter-particle spacing. The initial conditions at $z = 100$ are generated using the tool \textsc{N-GenIC} \citep{NGenic}, which is based on the Zel'dovich approximation (\citealt{zel1970gravitational}; see e.g. \citealt{White2014} for a comprehensive review). During the phase of linear evolution until $z \sim 100$, we do not take into account the effects of DMAF which would require modifying the initial conditions generator. The impact on our simulations is however expected to be small since e.g. \citet{Bertschinger2006} derives that early-time DMAF can enhance the abundance of small haloes with mass ranging from less than an earth mass to a few solar masses, which is many orders of magnitude below the scale which we can resolve in our simulations (each DM particle comprises a mass of $\sim7.95 \times 10^8 \ \text{M}_\odot$ in this simulation). 
\begin{figure*}
  \centering
  \noindent
  \resizebox{0.85\textwidth}{!}{
  \includegraphics[scale=1]{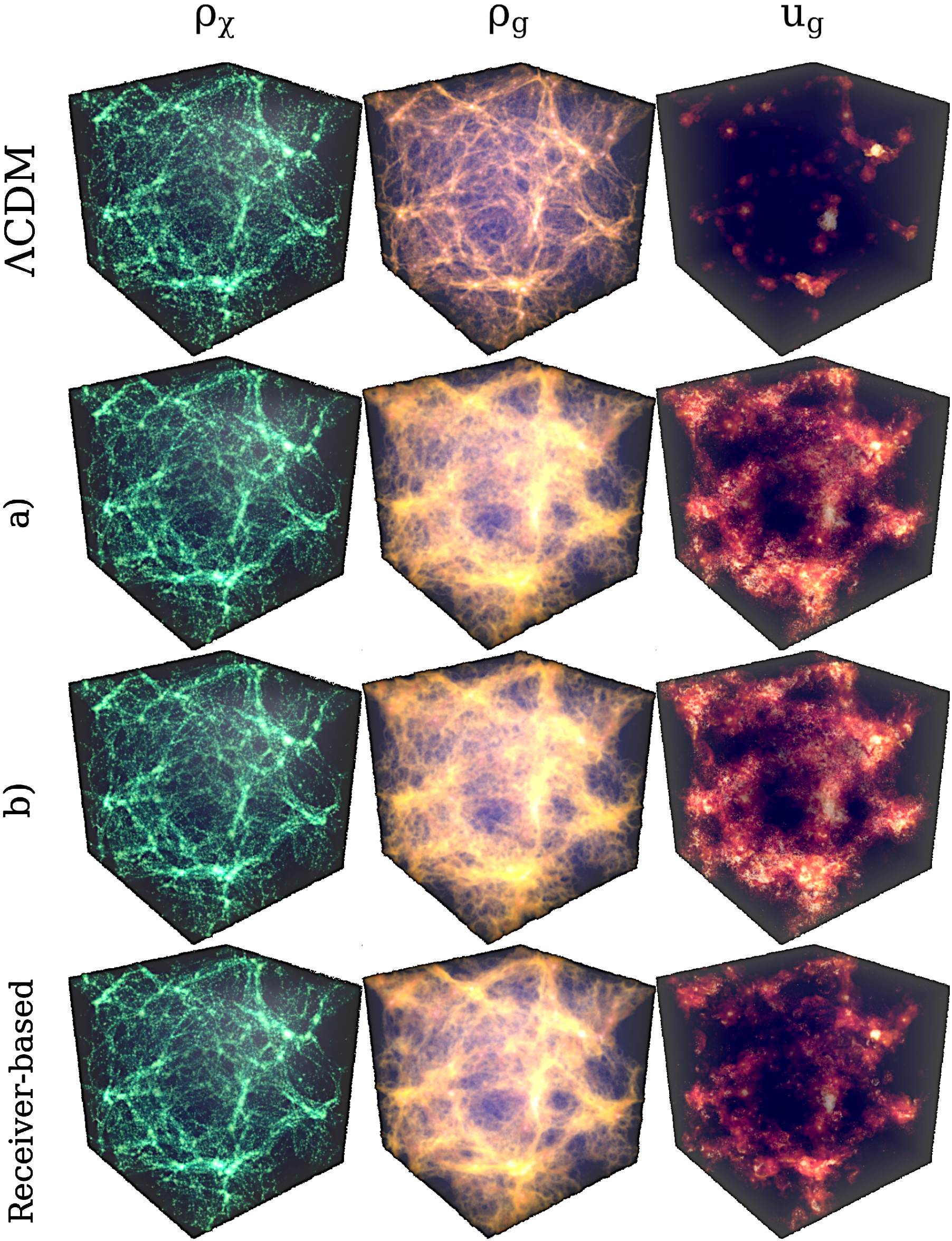}
  }%
  \caption{Results of the cosmological simulation with a light DM candidate of mass $m_\chi = 1 \ \text{MeV} \ c^{-2}$: DM density $\rho_\chi$ (\emph{left}), gas density $\rho_g$ (\emph{centre}), and specific internal energy of the gas $u_g$ (\emph{right}) at $z = 0$. The first row shows the fiducial $\Lambda\text{CDM}$ simulation without DMAF, the second and third row correspond to the donor-based method with choice of weights a) and b), respectively, and the last row shows the results of the receiver-based approach. DMAF evidently suppresses the formation of substructure -- most noticeable in the gas density plots.}%
\label{fig:Planck2015_result}
\end{figure*}
\par We compare the results for the choice of weights a) and b) 
with the receiver-based method; moreover, we run a fiducial $\Lambda \text{CDM}$ simulation \red{without DMAF}. Figure \ref{fig:Planck2015_result} shows a volume rendered representation (see \citealt{Garate2017}) of the DM density $\rho_{\chi}$, gas density $\rho_g$, and specific internal energy of the gas $u_g$. It is evident that the strong DMAF in this simulation suppresses the formation of substructure, and the gas density distribution is much more washed out than in the fiducial $\Lambda \text{CDM}$ simulation, as reported in \citet{Iwanus2017, Iwanus2019}. 
\par The weights $w_k$ only have a minor effect on the large-scale structure in this simulation; the density and energy distributions for case a) and b) strongly resemble each other. In contrast, the difference between the donor-based method and the receiver-based method manifests itself in the specific internal energy: for the receiver-based method, DMAF deposits less energy into the gas than for the donor-based method. This behaviour is in line with our findings in Subsection \ref{subsec:step_example}, where we showed that the receiver-based method may underestimate the DM density in case of a steep density gradient. As gas particles heat up and move away from the donating DM particles, this effect is likely to be magnified since the approximation for the DM density at the receiving gas particles deteriorates as the separation between donors and receivers increases. 
\begin{figure*}
  \centering
  \noindent
  \resizebox{\textwidth}{!}{
  \includegraphics{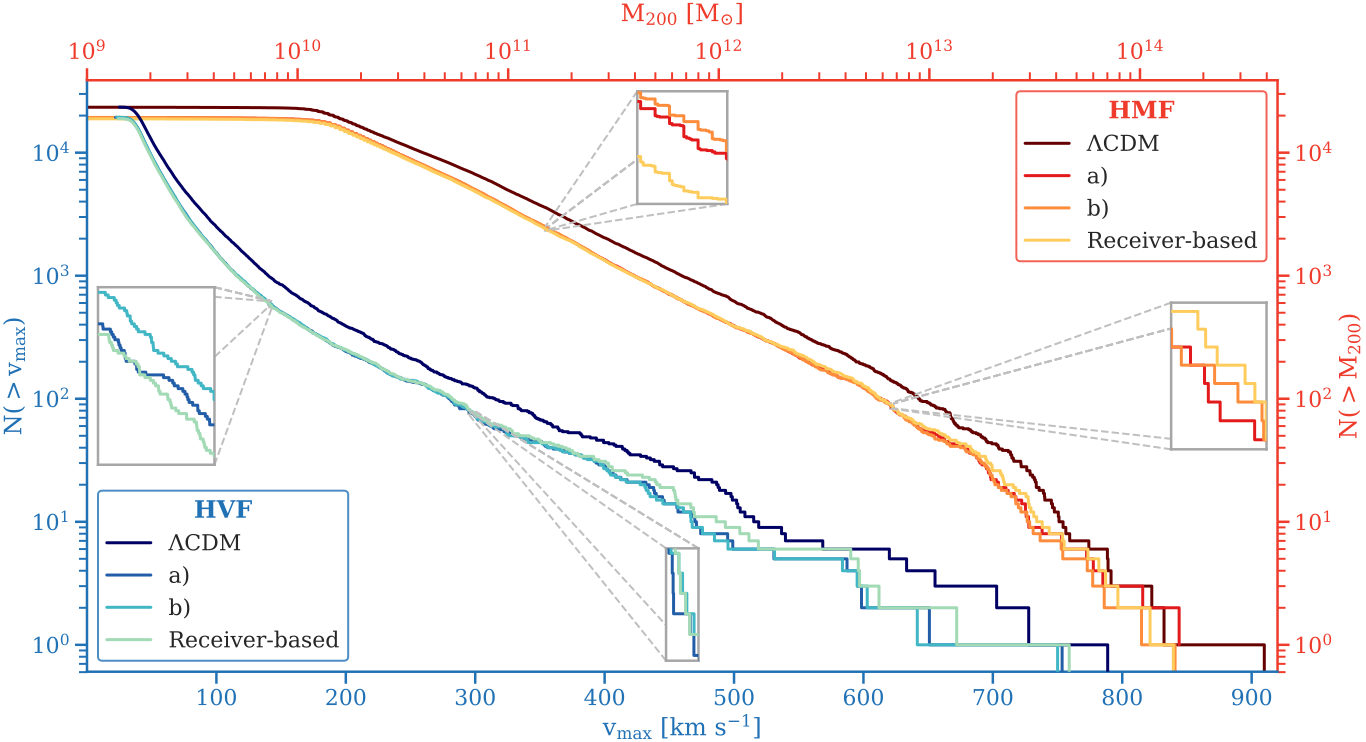}
  }
  \caption{Halo mass function (HMF) (\emph{red tones, upper-right corner}) and halo velocity function (HVF) (\emph{blue tones, lower-left corner}) for the cosmological simulation at $z = 0$. Inset plots show a zoom with 30-fold magnification. The DMAF quenches the formation of haloes of all sizes; additionally, it decreases the maximum halo velocities. The impact of the method for the choice of weights considered in this work on the haloes for which sufficient statistics are available ($N \gtrsim 30$) is small.}
\label{fig:Planck2015_HMF_HVF}
\end{figure*}
\par In order to compare the impact of the DMAF methods on the formation of structure, we calculate the halo mass function (HMF) and the halo maximum velocity function (HVF) for the different methods. The theoretical foundation of modelling the HMF has been laid by \citet{press1974formation} and it has since then evolved to a standard tool in the analysis of cosmological N-body simulations. Recent numerical studies investigating HMFs include \citet{Despali2016, Comparat2017, McClintock2018}.
\par The HVF is another important statistics that measures the gravitational potential within a halo \citep{Ascasibar2008} and is closely related to the baryonic content of the halo \citep{Comparat2017} in view of the well-established baryonic Tully--Fisher relation \citep{tully1977new, mcgaugh2000baryonic}. The authors of \citet{Ascasibar2008, Knebe2011, Onions2012} advocate the usage of the HVF as a metric for comparing (sub-)haloes due to its independence of the particular cut-off radius for the halo definition since the maximum halo velocity is typically reached within 20 per cent of the virial radius for moderately-sized haloes described by NFW profiles \citep{Muldrew2011}.
The maximum halo velocity is calculated as
\begin{equation}
v_\text{max} = \max_{r} \left\{\left(\frac{G M(r)}{r}\right)^{\nicefrac{1}{2}}\right\},
\end{equation}
where $M(r)$ is the mass enclosed within a radius of $r$.
\par Finding DM haloes is a highly non-trivial task and while different popular halo finders agree well on presence and location of the haloes \citep{Knebe2013a, Onions2012}, each halo finder leaves its imprint on the halo properties which can lead to differences of around 20 per cent \citep{Knebe2013, Onions2012} and impact the halo merger tree \citep{Avila2014a}. For a detailed overview on the zoo of halo finders, we refer to the references in this paragraph.
\par We opt for \textsc{VELOCIraptor} \citep{Elahi2019}, which is based on successive application of a friends-of-friends (FOF) algorithm in physical space and phase space. For the treatment of baryons, we select the ``DM+Baryons mode'' (see the reference for more details). 
\par Figure \ref{fig:Planck2015_HMF_HVF} shows the HMF and the HVF for the donor-based method with both choices for the weights, the receiver-based method, and the $\Lambda\text{CDM}$ simulation without DMAF. The mass $M_\text{200}$ denotes the virial mass for an overdensity parameter $\Delta_c = 200$ as usual. In this medium-resolution simulation, we are able to identify haloes of mass $\gtrsim 10^{10} \ \text{M}_\odot$. In total, we find 23446, 19361, 19401, 18904 haloes for $\Lambda\text{CDM}$, donor-based method a), donor-based method b), receiver-based method. Thus, DMAF curbs the formation of haloes at a level of $\sim 20$ per cent for this light DM candidate. This is because the DMAF energy heats up the gas which results in faster moving gas particles and an inhibited accretion of gas onto DM haloes. \red{The receiver-based approach produces slightly} fewer haloes than the donor-based approach. However, for haloes in the mass range $10^{10} - 10^{13} \ \text{M}_\odot$, where a statistically relevant number of haloes is available, the HMF and HVF for all methods agree very well with one another. In particular, the two choices of weights considered herein lead to similar results, as already suggested by Figure \ref{fig:Planck2015_result}. From the HVF, we infer that the DMAF energy inhibits the formation of haloes with deep gravitational wells and therefore with high maximum velocity. 
\begin{figure*}
  \centering
  \noindent
  \resizebox{\textwidth}{!}{
  \includegraphics{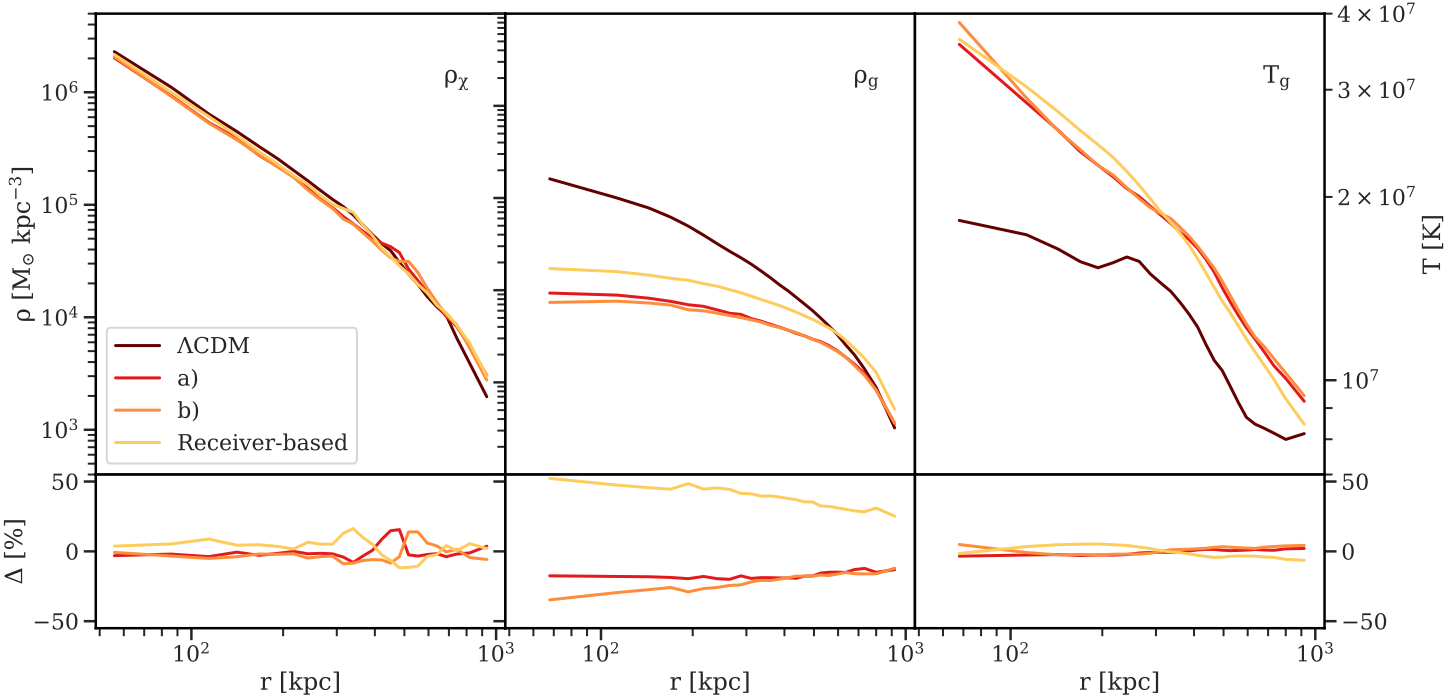}
  }
  \caption{\red{\emph{Largest galaxy cluster in the cosmological simulation}: radial plot of the DM density $\rho_\chi$ (\emph{left}), gas density $\rho_g$ (\emph{centre}), and gas temperature $T_g$ (\emph{right}) in a logarithmic scale. The lower panels show the relative difference towards the mean of a), b), and the receiver-based method. The different methods give a similar temperature profile; however, the gas density in the halo centre is more severely depleted with the donor-based method.}}
\label{fig:Halo_cosmo_plot}
\end{figure*}
\par \red{In Figure \ref{fig:Halo_cosmo_plot}, we have a closer look at the largest galaxy cluster in the cosmological simulation and plot the radial profiles of DM density $\rho_\chi$, gas density $\rho_g$, and temperature $T_g$. The haloes are matched across the different simulations by maximising $N_{a \cap b}^2 / (N_a N_b)$, where $N_{a \cap b}$ is the number of common DM particles in haloes $a$ and $b$, and $N_a$ and $N_b$ are the total numbers of particles in each halo. The galaxy cluster in the reference simulation without DMAF has a total mass of $1.4 \times 10^{14} \ \text{M}_\odot$. As expected, the DMAF has heated the cluster, in particular the inner region, and the different methods agree well with each other in terms of temperature. Interestingly, the gas density in the inner halo region is roughly $50$ per cent higher than average with the receiver-based method: as the heated gas particles move away from the halo centre and disperse to regions with lower DM density, the energy absorption is quickly reduced. In contrast, with the donor-based method, the DM particles in the halo centre continue depositing a large amount of energy into the gas particles even when they are receding from the centre, driving them further out.
However, this difference between the methods is moderate compared with the order of magnitude difference between the DMAF simulations and the reference simulation without DMAF. As for the two different weights, the gas density in the halo centre has decreased slightly more with the solid angle weighted injection. 
\par The fact that the difference in gas density between the two methods is not observed for the isolated halo (see Subsection \ref{subsec:halo_example}) suggests that the effects of the specific DMAF implementation become more relevant on larger time scales -- the galaxy cluster considered here is subject to DMAF for several Gyrs, compared with stronger DMAF for less than 100 Myrs in the case of the isolated halo.}
\par In this simulation, a very low mass has been taken for the DM candidate to showcase the impact of high DMAF energy rates on the large-scale structure of the Universe, although a WIMP particle this light has already been ruled out by various studies (e.g. \citealt{Leane2018}) assuming $2 \to 2$ s-wave annihilation. In a follow-up paper, more realistic scenarios will be considered and while the simulation herein lacks a model for gas cooling through effects such as bremsstrahlung, inverse Compton scattering, recombination, and reionization, these effects will be taken into account.

\section{Conclusions}
\label{sec:conclusions}
We have developed a novel numerical method for including DMAF in cosmological simulations. Our method can serve as a starting point for more elaborate dark sector models, e.g. DM annihilation into more general SM / dark sector products such as a dark radiation component. A varying degree of locality in the resulting deposition of the annihilation power can be modelled by weights accounting for heat generation through e.g. inverse Compton scattering, ionization, or excitation, and an extension of the weights presented herein will be addressed in future work. A careful treatment is required for synchronising two interacting species with different, individual time steps, namely injecting DM and receiving gas particles.
\par Our numerical results show good agreement with the receiver-based method presented in \citet{Iwanus2017} for simulations of an isolated halo and for a cosmological simulation, however, we present a toy example that showcases the conceptual differences between the two methods. \red{Over long periods of time, the donor-based method tends to reduce the gas density in halo centres somewhat more than the receiver-based method}. It is reassuring that for realistic test cases, numerical codes seem to be fairly robust with respect to the particular implementation of DMAF. 
\par Almost a century has passed since the first postulation of DM in the Universe and still, little is known about its nature. Joint efforts of experimentalists and theorists will be necessary for unravelling this mystery within the coming decades, and probing dark sector models by means of cosmological simulations can play a crucial role in this quest.

\section*{Acknowledgements}
The authors acknowledge the National Computational Infrastructure (NCI), which is supported by the Australian Government, and the University of Sydney HPC for providing services and computational resources on the supercomputers Raijin and Artemis, respectively, that have contributed to the research results reported within this paper. Thanks also go to Phil Hopkins and Volker Springel for making the codes \textsc{Gizmo} and \textsc{Gadget}-2 publicly available. F. L. is supported by the University of Sydney International Scholarship (USydIS).





\appendix

\section{Energy injection into an inactive gas particle}
\label{sec:inactive_gas}
\begin{figure}
  \centering
  \noindent
  \resizebox{\columnwidth}{!}{
  \includegraphics{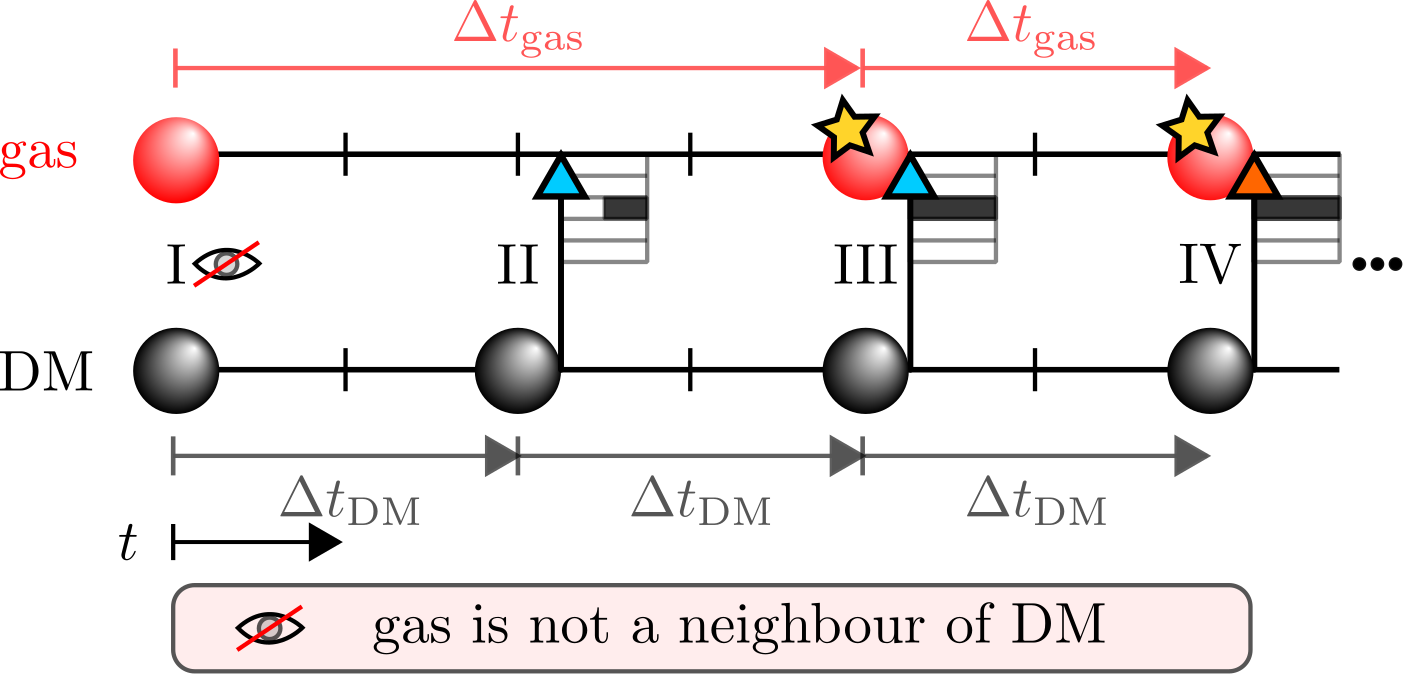}
  }%
  \caption[]{%
Time line of a DM and a gas particle: \\ %
I: Gas is not among the neighbours of DM and will receive no energy -- otherwise, $\Delta t_{\text{gas}}$ would be reduced. \\ %
II: End of DM time step, now, DM finds inactive gas particle and assigns DMAF energy rate to inactive gas. Moreover, $\Delta t_{\text{gas}}$ is restricted to be $\leq \Delta t_{\text{DM}}$ subsequently. \\ %
III: End of gas and DM time step, gas adds the energy set at II. DM informs gas about its time step again, gas reduces its time step. DMAF energy rate is updated and assigned to gas particle. \\ %
IV: End of gas and DM time step, gas adds the energy set at III. %
}%
\label{fig:Time_stepping_gas_inactive}
\end{figure}
We briefly discuss the case when a DM particle finds a receiver that is currently inactive. An exemplary time line for such a case is depicted in Figure \ref{fig:Time_stepping_gas_inactive}. A necessary requirement for this situation to occur is that the gas particle has not already been a neighbour at the beginning of its time step (at time I) -- otherwise, the DM particle would inform the gas about its smaller time step and the gas particle would subsequently reduce its time step. At time II, the DM particle finds the gas particle in the search for energy receivers. The DM particle assigns an energy rate to the gas particle and stores it in the bin corresponding to the DM time step. Note that although the DM time step is smaller than the gas time step, the gas particle does not zero out the energy rate prematurely since the zeroing out for each gas particle only takes place at the end of a gas time step when the gas particle is active. However, we must be careful to make sure that the energy rate is only applied between times II and III, that is over the second half of the gas time step. Therefore, the gas particle only stores a fraction of $\frac{dE_{i \to j}}{dt} \frac{\Delta t_\text{DM}}{\Delta t_\text{gas}}$, since all rates will be multiplied by its own time step $\Delta t_\text{gas}$, such that the amount of energy injected for the time interval between times II and III is $\frac{dE_{i \to j}}{dt} \Delta t_\text{DM}$, as it should be. This is illustrated by the half-filled array element.
\par In the exemplary case in Figure \ref{fig:Time_stepping_gas_inactive}, the gas particle is still a neighbour of the DM particle. This is however not necessary: even if the gas particle is no longer a neighbour of the DM particle at time III, it will inject the energy corresponding to the time interval between times II and III. Furthermore, the DM particle informs the gas particle at their first encounter at time II about its smaller time step; hence, the gas particle reduces its time step at time III, irrespective of it still being a neighbour of the DM particle or not.
\section{Cosmological expansion}
\label{sec:cosmological_expansion}
\begin{figure}
  \centering
  \noindent
  \resizebox{\columnwidth}{!}{
  \includegraphics{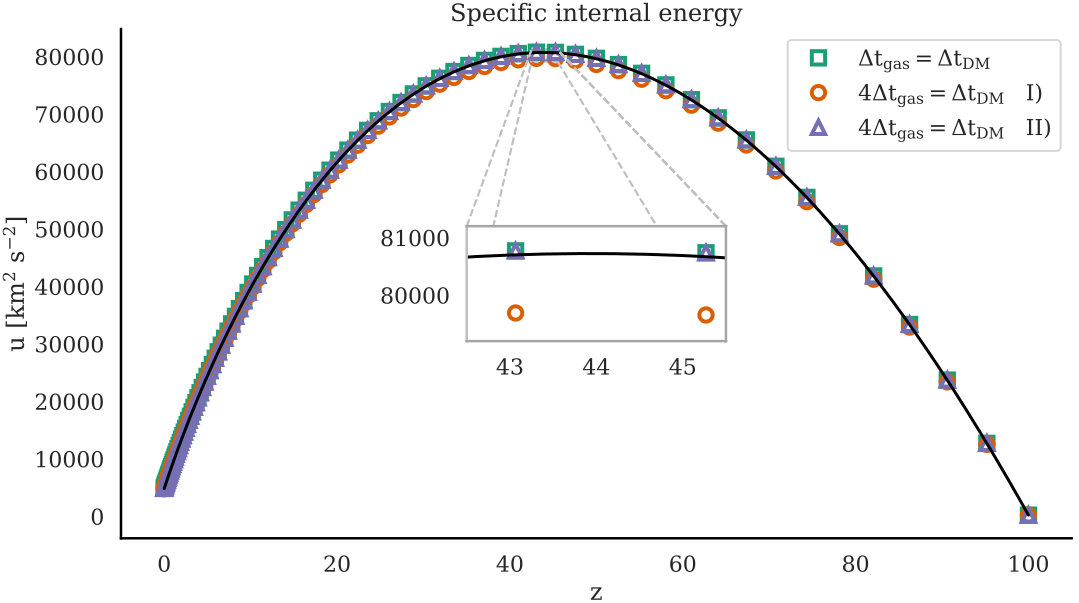}
  }%
  \caption{Numerical and analytical solution for a matter-dominated, homogeneous, isotropic universe and a DM candidate with energy $100 \ \text{keV}$. The numerical solution is plotted for the case of equal time steps for gas and DM and for the case of the DM time step being four times as large as the gas time step. In Variant I), the scale factor $a$ is evaluated at the end of the DM time step, which causes the energy injection to be slightly too small. In Variant II), the DMAF energy rate is adjusted to account for the current scale factor at the end of the gas time step.}
\label{fig:Homo_universe}
\end{figure}
For incorporating the expansion of the numerical universe in the simulation, the scale factor $a$ and Hubble parameter $H = \dot{a}/a$ are computed at each discrete time from the specified cosmological parameters. Since the density $\rho$ scales as $\rho = \rho_0 a^{-3}$, the DMAF energy rate in equation \eqref{eq:energy_i_j} scales with $a^{-3}$ as well. In order to deal with the cosmological expansion, we implemented two variants.
\par Variant I): if the DM mass loss due to annihilation is realised in the simulation, the total DMAF energy must be independent of the gas time steps, which might change while the injecting DM particle is inactive, in order to enforce $E = mc^2$ since the communication in our implementation is unilateral from the injecting DM particle to the receiving gas particle. For this reason, we evaluate $a$ at the end of the DM time step, which gives the correct value in case of $\Delta t_\text{gas} = \Delta t_\text{DM}$ and slightly underestimates the DMAF energy rate for $\Delta t_\text{gas} < \Delta t_\text{DM}$ due to the monotonicity of $a$. This bias converges to zero as the time steps become smaller.
\par Variant II): for a more accurate treatment of the cosmological expansion, we implemented a second option where we account for the Hubble flow of DM particles by rescaling the energy rate by a factor of $\left(a_\text{default} / a_\text{current} \right)^3$, where $a_\text{default}$ is the scale factor at the end of the injecting DM particle's time step and $a_\text{current}$ is the scale factor at the time when the gas particle adds the DMAF energy.
\par To illustrate the difference, we consider a homogeneous, isotropic universe without density fluctuations and evolve it from redshift $z_0 = 100$ to $z = 0$ using equal, fixed time steps for each species. For the DMAF deposition, we use the donor-based method with weights b). The analytical solution for the specific internal energy $u = u(z)$ is given by
\begin{equation}
u(z)=\frac{1+z}{1+z_{0}}\left[u_{0}+2 \kappa\left(1+z_{0}\right)^{\nicefrac{3}{2}}\left(1-\left(\frac{1+z}{1+z_{0}}\right)^{\nicefrac{1}{2}}\right)\right],
\end{equation}
where $\kappa = \frac{\langle\sigma v\rangle c^{2}}{m_{\chi}} \frac{\rho_\text{crit}}{H_{0}} \frac{\Omega_{\chi}^{2}}{\Omega_{b}}$ with Hubble constant $H_0$, critical density $\rho_\text{crit}$, and $u_0 = u(z_0)$ (see \citet{Iwanus2017}). Note that $a$ and $z$ are related by $1 + z = a^{-1}$. We take a DM candidate of mass $m_\chi = 100 \ \text{keV} \ c^{-2}$ with a thermal relic cross-section, $\Omega_b = 0.1573$, $\Omega_m = \Omega_b + \Omega_\chi = 1$, $\Omega_\Lambda = 0$, and $h = 0.6774$. We consider $16^3$ particles in a box with a side length of $50 \ \text{Mpc} \ h^{-1}$.
\par Figure \ref{fig:Homo_universe} shows the analytical solution and the numerical solution for three cases: equal time steps for DM and gas, four times larger time steps for the DM with variant I), and the same time steps with variant II). To be more precise, the time variable for cosmological simulations in \red{\textsc{Gizmo}} is $\log a$, and we fix $\Delta \log a_\text{gas} = 0.00225348$, $\Delta \log a_\text{DM} = \Delta \log a_\text{gas}$ for the first case, and $\Delta \log a_\text{DM} = 4 \ \Delta \log a_\text{gas}$ for the latter two cases.
\par For equal time steps and variant II), the numerical solution agrees very well with the analytical one, whereas the energy is slightly too low for variant I), as expected, for which reason we recommend to use variant II) if the DM mass loss from annihilation can be neglected.

\bsp	
\label{lastpage}
\end{document}